\documentclass[manuscript,screen,nonacm]{acmart}%

\setcopyright{none}

\usepackage{booktabs}
\usepackage{multirow}
\usepackage{caption}
\usepackage{subcaption}
\usepackage{graphicx}
\usepackage{amsmath}
\usepackage{amsfonts}
\usepackage{bbm}
\usepackage{dcolumn}
\usepackage{import}
\usepackage{pdflscape}
\usepackage{float}%
\usepackage{afterpage}
\usepackage{hyperref}
\usepackage{float}

\usepackage{enumitem}
\usepackage{xurl}

\usepackage[framemethod=tikz]{mdframed}
\usepackage{fontawesome}
\newcommand\coderepo{\url{https://github.com/xuanxuanxuan-git/hcxai}}
\newcommand\doi{\url{https://doi.org/10.1016/j.ijhcs.2024.103376}}

\definecolor{mycolor}{rgb}{0.122, 0.435, 0.698}
\definecolor{myblue}{rgb}{0.122, 0.435, 0.698}
\definecolor{mygreen}{rgb}{0.125, 0.525, 0.220}
\definecolor{myyellow}{rgb}{0.588, 0.439, 0.000}
\definecolor{myred}{rgb}{0.647, 0.114, 0.165}

\newmdenv[innerlinewidth=0.5pt,roundcorner=4pt,innerleftmargin=6pt,
          innerrightmargin=6pt,innertopmargin=6pt,innerbottommargin=6pt,
          linecolor=mycolor,backgroundcolor=mycolor!25!white]{mybox}
\newmdenv[innerlinewidth=0.5pt,roundcorner=4pt,innerleftmargin=6pt,
          innerrightmargin=6pt,innertopmargin=6pt,innerbottommargin=6pt,
          linecolor=myblue,backgroundcolor=myblue!25!white]{mybluebox}
\newmdenv[innerlinewidth=0.5pt,roundcorner=4pt,innerleftmargin=6pt,
          innerrightmargin=6pt,innertopmargin=6pt,innerbottommargin=6pt,
          linecolor=mygreen,backgroundcolor=mygreen!25!white]{mygreenbox}
\newmdenv[innerlinewidth=0.5pt,roundcorner=4pt,innerleftmargin=6pt,
          innerrightmargin=6pt,innertopmargin=6pt,innerbottommargin=6pt,
          linecolor=myyellow,backgroundcolor=myyellow!25!white]{myyellowbox}
\newmdenv[innerlinewidth=0.5pt,roundcorner=4pt,innerleftmargin=6pt,
          innerrightmargin=6pt,innertopmargin=6pt,innerbottommargin=6pt,
          linecolor=myred,backgroundcolor=myred!25!white]{myredbox}

\newlength\stextwidth

\begin{document}

\title[Comprehension Is a Double-Edged Sword]{Comprehension Is a Double-Edged Sword: \\Over-Interpreting Unspecified Information in Intelligible Machine Learning Explanations} 

\author{Yueqing Xuan}
\email{yueqing.xuan@student.rmit.edu.au}
\orcid{0000-0002-9365-8949}
\author{Edward Small}
\email{edward.small@student.rmit.edu.au}
\orcid{0000-0002-3368-1397}
\affiliation{%
  \institution{ARC Centre of Excellence for Automated Decision-Making and Society, School of Computing Technologies, RMIT University}
  \country{Australia}
}
\author{Kacper Sokol}
\email{kacper.sokol@inf.ethz.ch}
\affiliation{
\institution{Department of Computer Science, ETH Zurich}
\country{Switzerland}
}
\orcid{0000-0002-9869-5896}
\author{Danula Hettiachchi}
\email{danula.hettiachchi@rmit.edu.au}
\orcid{0000-0003-3875-5727}
\author{Mark Sanderson}
\email{mark.sanderson@rmit.edu.au}
\orcid{0000-0003-0487-9609}
\affiliation{%
  \institution{ARC Centre of Excellence for Automated Decision-Making and Society, School of Computing Technologies, RMIT University}
  \country{Australia}
}

\renewcommand{\shortauthors}{Xuan, et al.}

\begin{abstract}
Automated decision-making systems are becoming increasingly ubiquitous, which creates 
an immediate need for their interpretability and explainability. 
However, it remains unclear whether users know what insights an explanation offers and, more importantly, what information it lacks.
To answer this question we conducted an online study with 200 participants, which allowed us to 
assess explainees' ability to realise \textit{explicated information} -- i.e., factual insights conveyed by an explanation -- and \textit{unspecified information} -- i.e, insights that are not communicated by an explanation -- across four representative explanation types: model architecture, decision surface visualisation, counterfactual explainability and feature importance. 
Our findings uncover that highly comprehensible explanations, e.g., feature importance and decision surface visualisation, are exceptionally susceptible to misinterpretation since users tend to infer spurious information that is outside of the scope of these explanations. 
Additionally, while the users gauge their confidence accurately with respect to the information explicated by these explanations, they tend to be overconfident when misinterpreting the explanations. 
Our work demonstrates that human comprehension can be a double-edged sword since highly accessible explanations may convince users of their truthfulness while possibly leading to various misinterpretations at the same time. 
Machine learning explanations should therefore carefully navigate the complex relation between their full scope and limitations to maximise understanding and curb misinterpretation.
\end{abstract}


\maketitle

\vspace{-.2cm}%
\begin{mygreenbox}
\footnotesize%
\textbf{Highlights}%
\begin{itemize}[topsep=0pt,label=\faLightbulbO,leftmargin=.5cm,itemindent=0cm,labelwidth=.5cm,labelsep=0cm,align=left]%
    \item Users appear ignorant of explanations' limitations and tend to over-generalise factual insights.
    \item Users exhibit overconfidence when they misinterpret explanations, i.e., invent information that is not communicated by explanations.
    \item Highly comprehensible explanations are more likely to be misinterpreted by users.
    \item Easy-to-understand explanations, while highly comprehensible, tend to be misleading.
\end{itemize}
\end{mygreenbox}
\vspace{.2cm}%
\begin{mybluebox}
\footnotesize%
\begin{itemize}[topsep=0pt,leftmargin=1.7cm,itemindent=0cm,labelwidth=1.7cm,labelsep=0cm,align=left]
\item[\textbf{Source Code}]{\coderepo.}
\item[\textbf{DOI}]{\doi.} 
\item[\textbf{Cite Us}]{Yueqing Xuan, Edward Small, Kacper Sokol, Danula Hettiachchi, Mark Sanderson, Comprehension is a double-edged sword: Over-interpreting unspecified information in intelligible machine learning explanations, \emph{International Journal of Human-Computer Studies (IJHCS)}, 2024, 103376, ISSN 1071-5819, https://doi.org/10.1016/j.ijhcs.2024.103376.}
\end{itemize}
\end{mybluebox}

\section{Introduction}

Artificial Intelligence (AI), and Machine Learning (ML) in particular, can
deliver a range of benefits across the economy and society. These technologies
support diagnosis and early detection of dangerous health conditions in
hospitals~\citep{jiang2017artificial,davenport2019potential}, improve at-home
healthcare services~\citep{scale2022demand} and support personalised learning
and teaching~\citep{chassignol2018artificial,bhutoria2022personalized} among many other applications.
However, with automated decision-making tools becoming increasingly
sophisticated, and hence more opaque, we risk creating and using systems that we do
not fully understand or control. This situation not only has ethical
implications~\citep{dastin2022amazon}, but also raises concerns with respect to
accountability~\citep{kroll2015accountable}, safety~\citep{danks2017regulating}
and liability~\citep{kingston2016artificial}. 

Researchers and regulatory bodies have urged providers of AI tools to explain
predictions and decisions output by these systems to the public, especially the affected individuals~\citep{voigt2017eu,guidotti2018survey,aida2023}. The European Union's General Data Protection Regulation (GDPR) proposed a right
to explanation, which stipulates that all individuals should be able to obtain
``meaningful explanations of the logic involved'' in automated
decision-making~\citep{voigt2017eu}. The Canadian Artificial Intelligence and
Data Act (AIDA) requires high-impact AI systems to be transparent by providing
information that is sufficient to allow the public to understand their
capabilities, limitations and potential impact~\citep{aida2023}. However, it
remains unclear what constitutes a ``meaningful''
explanation~\citep{goodman2017european}. To complicate matters further,
explanations of AI systems can be deceptive, e.g.,~they can be manipulated to
intentionally mislead the
public~\citep{eiband2019impact,lakkaraju2020fool,danry2022deceptive,sokol2024interpretable}. 
While previous work has evaluated whether different machine learning
explanations are understandable to diverse
stakeholders~\citep{cheng2019explaining,lage2019human,kaur2020interpreting,bove2023investigating},
research on whether users are aware of the limitations of the information
explicated by ML explanations is still largely missing. 

The availability of AI explainability tools has soared in recent years. Popular
explanatory mechanisms include deploying inherently transparent models such as
tree-based predictors~\citep{rudin2019stop}
or generating post-hoc explanations of black-box
models~\citep{koh2017understanding,wachter2017counterfactual,guidotti2018survey}. 
While being essential for high stakes domains~\citep{rudin2019stop},
transparent models may not necessarily engender understanding, especially in
lay
audiences~\citep{abdul2018trends,miller2019explanation,sokol2023reasonable}, 
thus appropriate explanatory mechanisms may need to be deployed to facilitate user comprehension.
Genuine understanding of an explanation requires the users not only to internalise the information that it offers but also to appreciate the information that remains unspecified. 
The latter aspect concerns over-generalisation of explanations beyond their scope, which 
is of particular importance in view of the illusion of explanatory depth, i.e.,~people's limited knowledge and incorrect perception 
that they understand a topic with far greater depth than they actually
do~\citep{rozenblit2002misunderstood}. 

In this paper, we study user comprehension of \emph{information} that is \emph{explicated} and that remains \emph{unspecified} by selected machine learning explanations. The former is understood as factual insights that are provided in an explanation; the latter encompasses information that the explanation is not designed to communicate but may be misconstrued by the users. In practice, information about ML models is often selectively presented in explanations to maximise comprehensibility; however, the line separating the explicated from the unspecified information is often blurry, which may unintentionally contribute to users over-generalising their interpretation of the explanations. Consequently, users can internalise invalid explanatory insights even when neither deception nor malice were intended. We refer to such explanations -- for which unspecified information can be easily misconstrued and over-interpreted -- as \emph{misleading}.

As an example, consider counterfactual explanations, which inform users about
the smallest possible change to a feature vector that results in a desirable
outcome~\citep{wachter2017counterfactual}. While they report a subset of
features whose values need to be changed, they do not communicate the (local or
global) importance of these features for the model's decision. 
Nonetheless, such an explanation can mislead some users if they misunderstand its message and incorrectly infer insights about feature importance, which information is not communicated, i.e.,~remains unspecified. 
In such a case, the explanation could potentially be harmful since it is
intrinsically faithful and trustworthy, which is likely to convince a user of
its truthfulness and utility~\citep{jacovi2021formalizing}, while at the same
time possibly leading to various misinterpretations.

Current research has explored other types of \emph{misleading} explanations. 
\citet{lakkaraju2020fool} manipulated the explanation generation process to
create insights that misrepresent black-box models, thus disguise their lack
of fairness. The authors then evaluated whether these unfaithful explanations can
trick users into believing that such black-box models are, in fact, fair. Our
paper differs from their work as we generate genuine and faithful explanations
without any intention to deceive the users. We also focus on misinterpretation
of unspecified information, which is a different facet of explanations being
misleading. 
\citet{eiband2019impact} explored the impact of \emph{placebic} explanations --
which are constructed artificially and provide no useful information -- on user
trust. Our work differs from their approach as we use meaningful explanations
that are popular in the literature. Our work is complementary to the
aforementioned papers since users can have high trust in unfaithful
explanations regardless of their level of comprehension, and highly
comprehensible explanations do not necessarily guarantee user trust.

In this work, we designed and executed an online user study with 200 participants to systematically investigate four representative explainability approaches: model architecture, decision surface visualisation, counterfactual explainability and feature importance. Each explanation is generated for two inherently transparent predictive models: logistic regression and decision tree (we justify our study design choices in Section~\ref{sec:design}). 
Specifically, to assess user comprehension of both explicated and unspecified information, we designed two types of comprehension statements for each explanation: a \emph{statement of explicated information}, which assesses whether users can understand the insights communicated by an explanation; and a \emph{statement of unspecified information}, which gauges whether users are aware that certain information remains unspecified by an explanation, thus can appreciate its limitations. 

In our study, the participants were asked to judge these two statements for each explanation of one of the ML models. We also asked the participants to report their confidence in their judgement as well as their perception of whether an explanation was easy to understand and sufficiently detailed. We further collected demographic information. With this set-up (the details of which are described in Section~\ref{sec:user-study}), we seek to answer the following research questions:
\begin{description}
    \item [RQ1] How do user comprehension of explicated information and, more importantly, misinterpretation of information that remains unspecified differ across diverse machine learning explanations?
    \item [RQ2] How do user confidence in their (mis)interpretation, their perception of explanation difficulty and their appreciation of explanation information richness differ across explanations?
    \item [RQ3] How do individual characteristics -- such as education level, algorithm literacy, technical background, graph literacy and experience with explainable AI user studies -- affect user (mis)interpretation of explanations?
\end{description}

We find that it was easier for our participants to identify explicated information when compared to unspecified information across all four explanation types. 
Additionally, when the explicated information offered by an explanation was highly comprehensible, 
the users were easily misled as they tended to more readily misconstrue unspecified information, thus misinterpret the explanation. 
For example, the participants were significantly more likely to understand the information explicated by feature importance and decision surface visualisation in comparison to the other two explanation types, but these two explanation types were also more misleading. 
Counterfactual explainability and model architecture were less intelligible in terms of explicated information, but also less misleading. 
Furthermore, the participants who identified \emph{explicated information} correctly reported higher confidence in their interpretation than their peers who did not; 
for \emph{unspecified information}, the participants who understood the limitations of the explanations were less certain than their overconfident peers who misinterpreted the explanations. Full study results are presented in Section~\ref{sec:results}, followed by an in-depth discussion in Section~\ref{sec:discussion} and a conclusion in Section~\ref{sec:conclusion}. 

In summary, our work demonstrates that comprehension of the information explicated by ML explanations can be a double-edged sword as it may prompt users to misconstrue explanatory insights and be overconfident in their incorrect beliefs. 
Specifically, the contributions of our work are three-fold.
\begin{enumerate}
    \item We introduce the novel concept of \emph{unspecified explanatory information}, which is misconstrued by explainees based on a sound and faithful ML explanation. We then highlight the importance of assessing 
    explicated and unspecified information in explainable AI -- i.e.,~the scope of explanatory insights and the (implicit) limitations thereof -- which can lead to their unintended misinterpretation. 
    We demonstrate this phenomenon through a comprehensive user study based on a collection of representative machine learning explanations that are faithful to the underlying predictive model, thus ensuring the ecological validity of our study. 
    \item We demonstrate that comprehension is a double-edged sword: an explanation can at the same time be highly comprehensible and susceptible to misinterpretation as users are less attuned to the limitations of ``easy-to-understand'' explanations. 
    \item We design a highly flexible and reusable framework for evaluating whether an explanation is comprehensible in view of its explicated and unspecified information, thus the degree to which an explanation can mislead explainees. This contribution enables others to evaluate different types of explanations situated in distinct domains, allowing them to identify and curb any form of their misinterpretation, e.g.,~by explicitly indicating their limitations and introducing complementary explanations (to the same effect).
\end{enumerate}

\section{Related Work}

Evaluating the intelligibility of different ante-hoc interpretability and post-hoc explainability approaches across diverse stakeholders is a popular research topic. 
This paper builds upon prior work in eXplainable AI (XAI) and human comprehensibility of such techniques, a review of which follows. 

\subsection{Explainable AI}
There has been increasing interest in safety-critical industries to leverage
machine learning for high stakes predictions. Many ML
models are constructed as black boxes whose internals are either unknown to the
observer or impossible to interpret by humans~\citep{rudin2019stop}. 
A range of XAI tools are commonly used to explain the logic of automated
decision-making. For example, transparent models that are self-explanatory are
a form of ante-hoc interpretability in which the model itself constitutes an
explanation~\citep{sokol2023reasonable}. Post-hoc explanations -- 
defined as an ``interface'' between humans and a predictor that is both
comprehensible and an accurate proxy of the
underlying predictor~\citep{guidotti2018survey} -- 
provide a different mechanism to inspect the behaviour of a data-driven model.

For example, a \emph{post-hoc counterfactual} explainer
familiarises humans with unknown model behaviour by simulating
some hypothetical (input) circumstances under which the output changes to the desired state, aiming to develop user trust~\citep{del2024generating}.
Such insights, however, may not be reliable and truthful with respect to the
underlying predictor even if they are of high or full
fidelity~\citep{rudin2019stop,sokol2020limetree}. For
example, \citet{aivodji2019fairwashing} and \citet{lakkaraju2020fool}
manipulated explanations of black-box models to hide their unfair
decision-making by approximating them with seemingly fair transparent models
from which explanations were derived. In such cases, if users misunderstand a
model after receiving an explanation, it is unclear 
whether this is caused by misinterpretation of a post-hoc explanation or its
distorted representation of the underlying model~\citep{small2023helpful}. 
In this work, we take a step back from sophisticated models and their post-hoc explainability, and instead look at \emph{inherently transparent} models and their ante-hoc interpretability, which is guaranteed to be truthful and faithful. 

Additionally, explanations differ in their scope -- global or local -- and the
type of information they communicate~\citep{sokol2020explainability}.
Global explanations describe the overall behaviour of a machine learning model; local explanations pertain to individual predictions.
An inherently interpretable model constitutes a global explanation since we are
able to understand the logic of an entire model, nonetheless it also offers
local insights when following its reasoning for individual
instances~\citep{guidotti2018survey}. 
Feature-based explanations, e.g.,~feature importance or influence, show how much a feature impacts a prediction (locally) or how important it is for the model as a whole (globally) or its individual prediction (locally).
As a result, such explanations differ in their general strengths, weaknesses and applicability to
real-world use cases~\citep{retzlaff2024post}.
Given that multiple explanatory tools of varying scope are developed to help users understand predictive models from different perspectives, it is crucial to investigate the effectiveness of these explanations. 
This does not only require the explainees to correctly identify the scope, but also to be aware of the limitations of an explanation. In this paper, we explore both of these perspectives with a strong focus on whether users can 
recognise what information an explanation lacks, i.e.,~its limitations, to avoid their over-generalisation.

\subsection{User Comprehension}%

Explanations are intended to assist users in understanding the general functioning as well as selected details of an ML model from varying perspectives. 
Because of the breadth and scope of these objectives current literature lacks a
consensus on what actually constitutes user
comprehension~\citep{meske2022explainable,sokol2021explainability}. 
\citet{cheng2019explaining} considered user understanding in terms of whether
explainees can pick up the influence of features on a model's output and
simulate predictions of that model given feature changes. 
\citet{bove2022contextualization} captured user understanding from the
perspective of identifying feature influence on a single instance and the scope
of an explanation. 
In another work, which explored the intelligibility of counterfactual
explanations, \citet{bove2023investigating} assessed whether users can identify
that the provided information was a counterfactual. 

In addition to post-hoc explainability, researchers have also explored the
comprehensibility of inherently transparent models, i.e.,~ante-hoc
interpretability~\citep{adadi2018peeking}. 
\citet{huysmans2011empirical} explored the comprehensibility of a number of
such models by considering whether users can derive the model's output and
identify feature influence. 
Similarly, \citet{bell2022s} measured interpretability as the users' ability to
anticipate the output of an ML model or identify its most important feature. In
another work, \citet{lage2019human} evaluated how well users can simulate the
predictions of different transparent models. The authors asked participants to
derive and verify outputs as well as anticipate changes to outputs based on
alterations to input values.

User-based evaluation of explainability and interpretability can be premised on
objective or subjective assessment of the explainees' understanding of predictive
systems~\citep{sokol2024what}. 
The former allows for a quantitative approach that employs a questionnaire
consisting of a collection of curated questions about a predictive system that
are aligned with a selected definition of
comprehension~\citep{cheng2019explaining,bove2022contextualization,bove2023investigating}. 
Subjective understanding, on the other hand, is often based on the Explanation
Satisfaction Scale~\citep{hoffman2018metrics}, which asks users whether an
explanation is understandable and fulfils their needs, among
others~\citep{shin2021effects,bove2022contextualization,bove2023investigating}. 
Our user study builds upon both of these paradigms and extends them beyond evaluating the comprehension of information explicated by an explanation to also measure whether users can recognise (often implicit) explanation limitations, thus avoid misconstruing information that remains unspecified by the explanation. 
In this paper, we employ a questionnaire to evaluate objective understanding and subjective perception of explanations, accounting for both of the aforementioned perspectives. 

A related line of research has studied \emph{misleading} explanations, but the definition of this property is inconsistent. 
\citet{eiband2019impact} explored whether users would misplace trust in
explanations that contain void information; their main focus was on meaningless
and invalid explanations.
\citet{lakkaraju2020fool} assumed misleading
explanations to be those that deliberately misrepresent black-box models to
disguise their unfairness; 
the authors created misleading explanations  
so that their recipient cannot uncover the underlying bias even if they interpret those explanations correctly. 
In contrast, 
we experiment with well-established (ante-hoc) interpretability techniques that are designed to maximise intelligibility of (inherently transparent) predictive models, focusing on over-interpretation of \emph{unspecified} information.  

Our conception of how users misinterpret explanations is also closely related to the
folk concept of behaviour~\citep{malle1997folk} and the illusion of explanatory
depth~\citep{rozenblit2002misunderstood}.
\citet{jacovi2021formalizing} attributed the failure where user understanding differs from what the
explanation attempts to communicate to folk concepts of behaviour. 
Furthermore, \citet{chromik2021think} examined if users fell for an illusion of
explanatory depth when interpreting additive local explanations. However, both
pieces of work focused exclusively on explanations based on local feature
attribution,
and observed that users inferred global information from local explanations. Our work makes use of a comprehensive set of explanations, including both local and global explanations, and examines users' two-way over-interpretation: inferring local from global and vice versa.
We also find that users infer excessive amount of
local information that is not intended to be communicated in a local
explanation, with our results being complementary to those reported
by \citet{chromik2021think}.
To the best of our knowledge, we present the first comprehensive user study that identifies what information is missing from representative explanations, and evaluates users' comprehension of information that remains unspecified by these explanations.

\subsection{Explanation Effectiveness}

Extensive research has been done to evaluate the effectiveness of ML explanations in improving user understanding.
\citet{cheng2019explaining} studied the ability of different explanation
interfaces to increase people's understanding of a data-driven university
admissions system. They concluded that interactive and ante-hoc explanations
increase participants' objective understanding of the algorithm, whereas their
subjective understanding does not increase for the latter aspect of explainability,
possibly due to information overload.
\citet{bell2022s} compared two interpretable models -- linear regression and
decision tree -- with a black-box model applied to public policy domains. Their
results indicate that black-box models can be perceived just as explainable as
inherently interpretable models, possibly due to user confusion
caused by the overwhelming amount of information offered by interpretable models.  
\citet{bove2023investigating} found that plural counterfactual examples
increase objective understanding and satisfaction of explainees. In another
work, \citet{bove2022contextualization} showed that augmenting local feature
importance explanations with contextual information improves their subjective
understanding. 

Some early XAI work focused exclusively on interpretable models and explanations thereof. \citet{huysmans2011empirical} empirically compared the comprehensibility of
predictive models based on decision tables, trees and rules, and demonstrated
that decision tables provide a significant advantage in terms of
comprehensibility. \citet{lim2009and} investigated which of ``Why?'', ``Why
not?'', ``How to?'' and ``What if?'' explanations are more effective in helping
users understand a decision tree model. Their results showed that ``Why?''
explanations yield the best understanding.
Despite a wealth of literature on the comprehensibility of different explanations, little effort has been made to compare local and global explanations, or to explore whether users can identify their limitations. Our work fills this gap by focusing on both local and global interpretability of two inherently transparent predictive models, thus guaranteeing the correctness of these insights. 

\begin{table}%
    \centering
\small
    \caption{Overview of four different explanations for two predictive models used in our study: logistic regression and decision tree.
    In the experiment, these explanations are paired with an informative description as outlined in Table~\ref{tab:lr_questions} and Table~\ref{tab:dt_questions} in Appendix~\ref{app:exp}.
    The logistic regression model is displayed with specific weights and accompanied by a coefficient table to assist the users in interpreting the equation.
    \label{tab:explainer}}
    \begin{tabular}{@{}p{.4cm}wc{4cm}wc{3cm}p{2cm}p{4cm}@{}}
    \toprule
    & Feature importance & Decision surface & Counterfactual & \multicolumn{1}{c}{Model architecture} \\
    \midrule
    \parbox[t]{2mm}{\multirow{1}{*}{\rotatebox[origin=c]{90}{Log.\ Reg.}}} &
        \raisebox{-0.75\height}{\includegraphics[width=4cm]{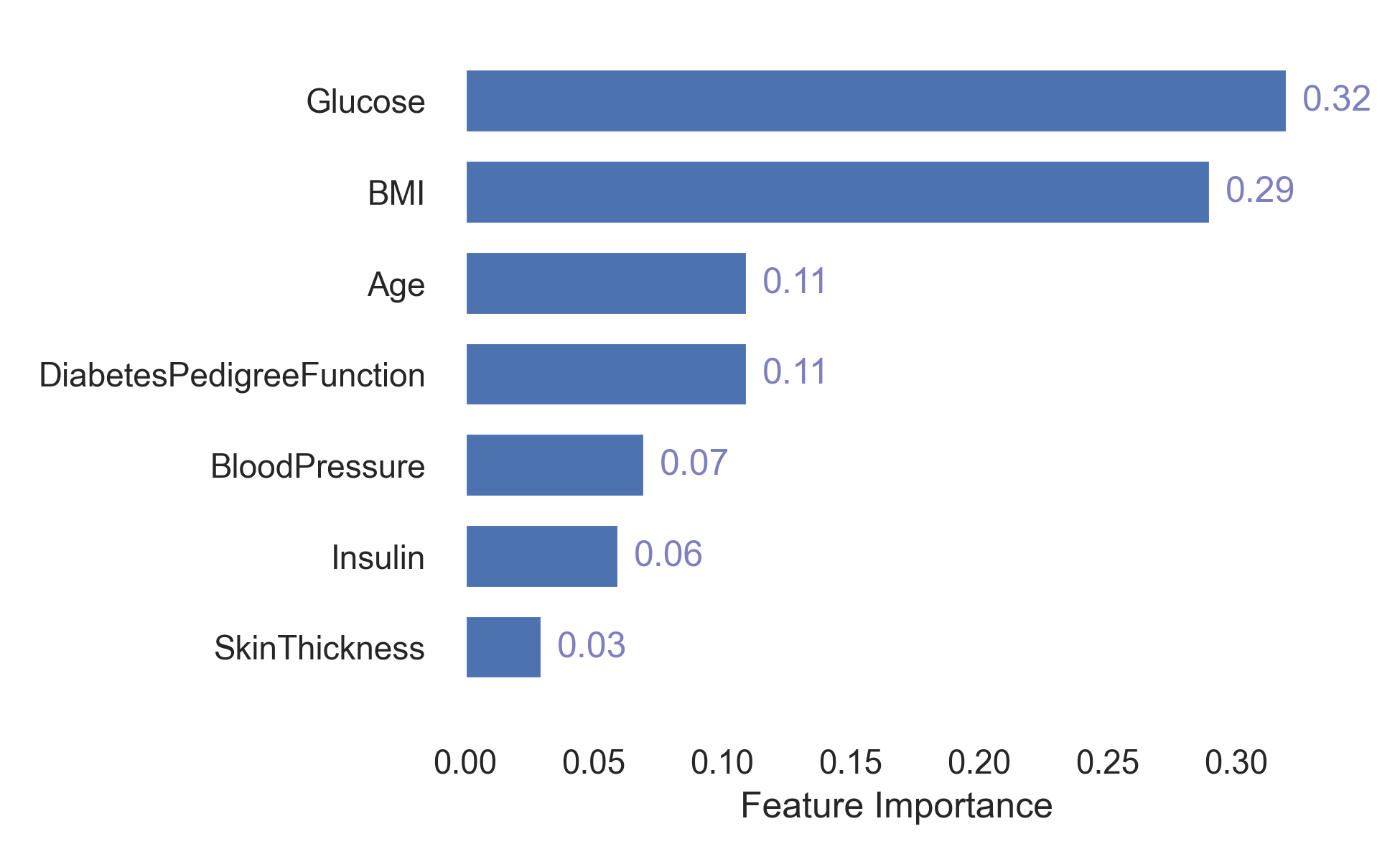}} &
        \raisebox{-0.75\height}{\includegraphics[width=3cm]{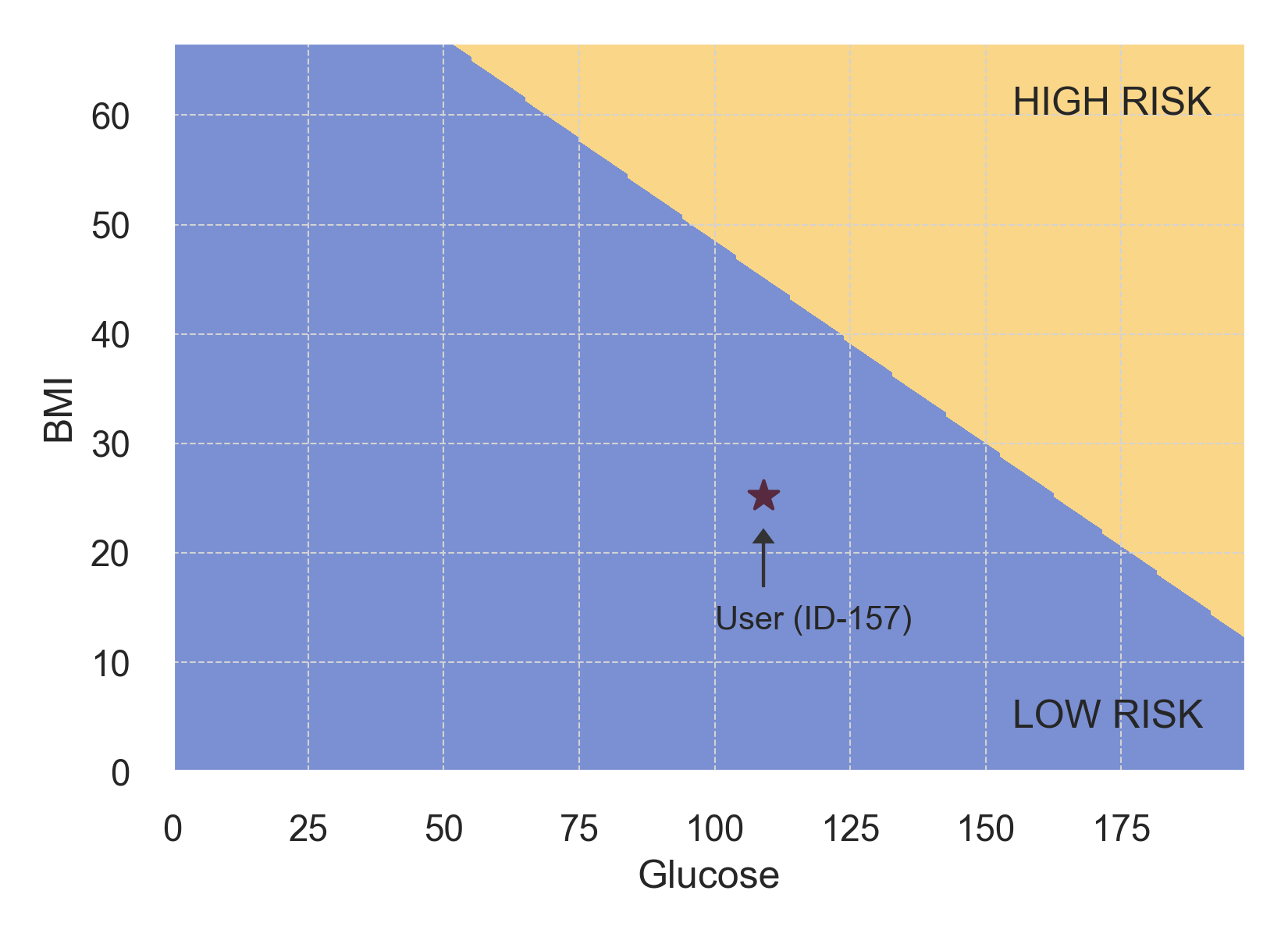}} &
        {\em\small Had your Glucose been 150 and BMI been 30, you would have been predicted with low risk.}  &
        {\small
        $ z(x) = \omega_0 + \omega_1 x_1 + \cdots + \omega_7 x_7 $
        
        $ f(z) = \Large \frac 1 {1 + e^{-z}} $
        
        $\text{Result} = \begin{cases} \text{high} \quad \text{if } f(z) \geq 0.5\\ \text{low} \quad\; \text{otherwise} 
            \end{cases}$}
        \\%
    \parbox[t]{2mm}{\multirow{1}{*}{\rotatebox[origin=c]{90}{Decision Tree}}} &
            \raisebox{-0.75\height}{\includegraphics[width=4cm]{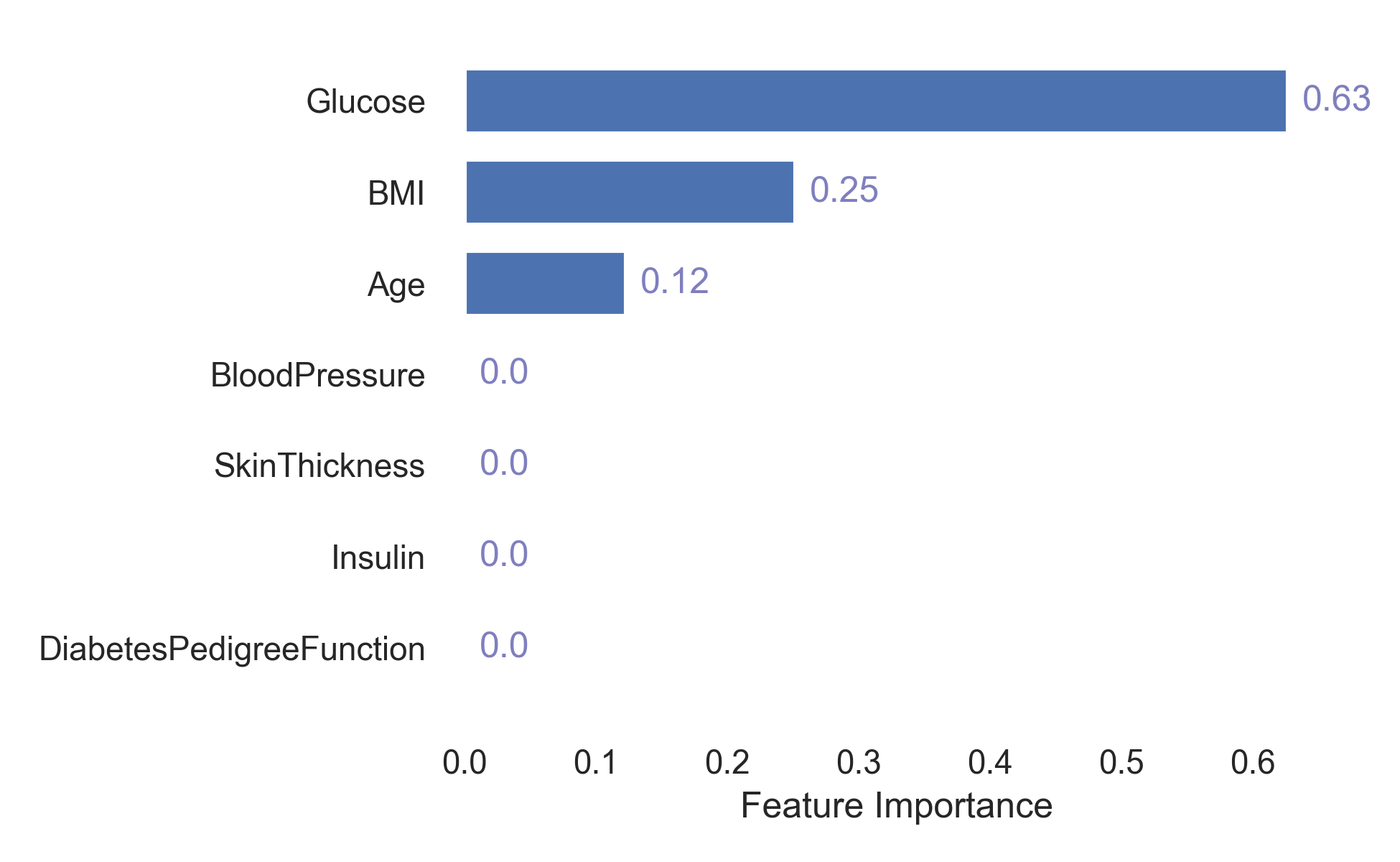}} %
            &
            \raisebox{-0.75\height}{\includegraphics[width=3cm]{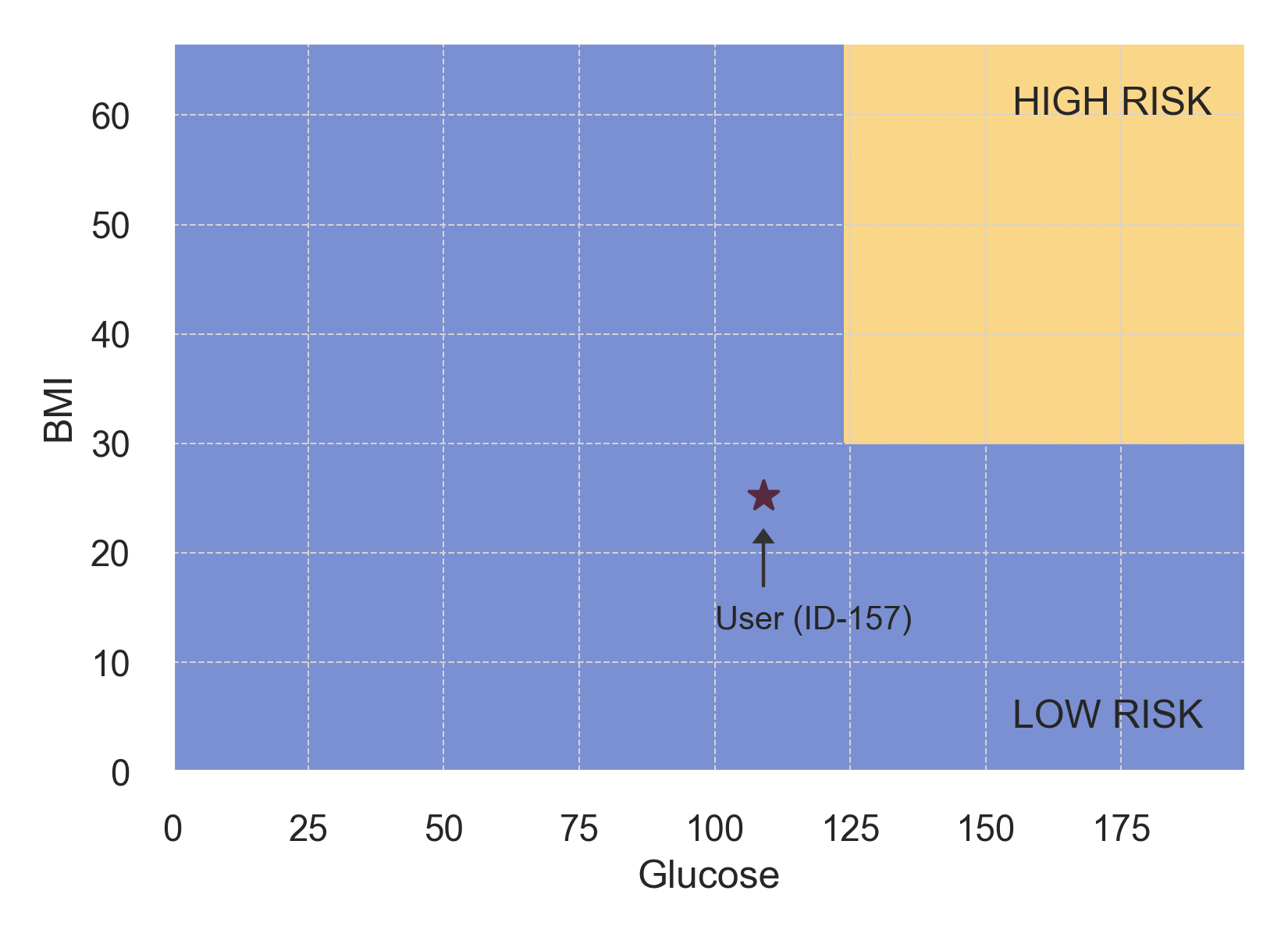}} &
            {\em\small Had your Glucose been 150 and BMI been 29, you would have been predicted with low risk.} &
            \raisebox{-0.75\height}{\includegraphics[width=4cm]{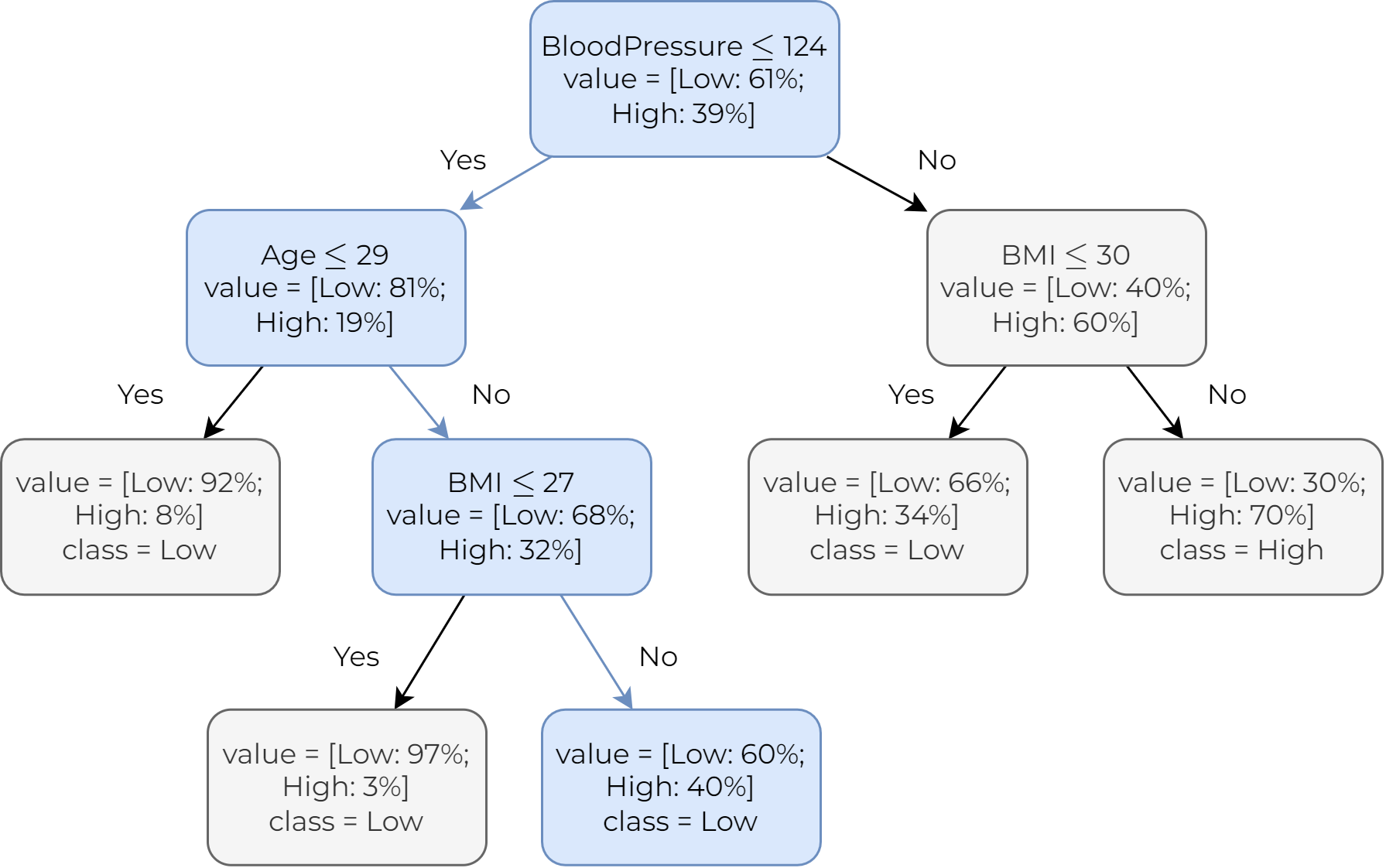}} \\
    \bottomrule
\end{tabular}

\end{table}

\section{Study Design}\label{sec:design}%
Before reporting our results, we discuss the choice of machine learning models and their interpretability techniques as well as use case scenarios in our experiments. 

\subsection{Predictive Models}\label{sec:ml-model}%

In this study, we explore the intelligibility of explanations for
\emph{decision tree} and \emph{logistic regression} models. We select these two
ML models because they are inherently transparent and each of their parameters
is directly interpretable, therefore we can generate a diverse range of
faithful and reliable ante-hoc explanations in addition to using the model
itself as an explanatory
artefact~\citep{guidotti2018survey,jakka2023explainable}. 
These two models are commonly used as explanatory artefacts given their structural simplicity and because they provide the most comprehensive (full) information about the model's functioning. 
Additionally, these predictive models are commonly used in high stakes domains
because of their transparency and
interpretability~\citep{rudin2019stop,lundberg2020local}. 
Both models are capable of predicting binary outcomes, can be trained as
to maintain low complexity, and tend to achieve comparable predictive
performance after appropriate hyper-parameter tuning, yet they operate in a
distinct manner~\citep{guidotti2018survey}. 
For our experiments, we use scikit-learn~\citep{pedregosa2011scikit} to train
the decision tree 
and logistic regression 
models, and perform hyper-parameter tuning to ensure their high predictive performance.

We choose transparent ML models to ensure perfect fidelity of their
explanations, i.e.,~the explanations are guaranteed to accurately reflect the
internals of the underlying models. As such, we are able to explicitly verify the quality of
the generated explanations, thus control for the confounding factor where a
user misunderstands an explanation because it lacks fidelity rather than
intelligibility. In contrast, post-hoc explanations of black-box models are
generated by only approximating their behaviour, which can result in unfaithful
explanations that fail to precisely capture the functioning of the explained
model~\citep{du2019techniques}. 
Given the opaqueness of black-box models, it is technically challenging to
design their explanations with perfect fidelity; we are thus unable to control
for the aforementioned confounding factor. In addition, explanations of
black-box models can be easily manipulated due to the nature of their
generation process -- they only approximate the black
box~\citep{lakkaraju2020fool,sokol2024interpretable}. Contrarily, ante-hoc
interpretations of transparent models cannot be easily doctored, which ensures
that user misunderstanding can be easily attributed to a concrete part of the
explanatory pipeline.

\subsection{Explainers}\label{sec:explanations}%

We choose four explanation types -- \textit{model architecture}, \textit{decision surface visualisation}, \textit{feature importance} and \textit{counterfactual explainability} -- that are commonly used to assist humans in understanding ML models. Each of these explanations can be appealing to a different stakeholder regardless of their level of expertise. 
Table~\ref{tab:explainer} shows examples of the explanations used in our user study for the two ML models. 
These explanations differ in their scope, hence the amount of information they communicate. 
We use feature importance as a global explanation that conveys how important each feature is for the overall performance of an ML model. 
We employ decision surface visualisation to display the local behaviour of an ML model, i.e.,~its change in prediction, when varying \emph{two} selected features for a single data point while the other features remain unchanged, which allows us to plot this explanation in two dimensions. 
Given the difficulty of visualising a high-dimensional decision surface in an explanation (seven dimensions for our dataset), it is necessary to explicitly communicate that the remaining features are kept constant to enable the generation of this two-dimensional explanation. This assumption is explicitly stated in the description of the decision surface explainer to ensure that the explainee can clearly understand what is being shown.
Counterfactual explainability communicates the local behaviour of a model for a
given data point by informing the explainee about the smallest change to the
feature vector that results in a desirable
outcome~\citep{wachter2017counterfactual}.
The architecture of transparent models is often used as an explanation itself
and allows the users to simulate their predictive behaviour (local) or grasp their
overall mechanics (global)~\citep{lim2009and,huysmans2011empirical,bell2022s}. 

To ensure the (ecological) validity of our findings, we use existing XAI tools to
generate these explanations. Feature importance for logistic regression is
based on the absolute value of the model's coefficients; for decision trees, it
is based on the decrease in Gini impurity calculated for the training
data~\citep{breiman2017classification}. 
To unify the representation of feature importance across these two models, we normalise the importance values such that they sum up to $1$ and sort feature importance in descending order.
To visualise the decision surface we randomly select two features with non-zero importance and display how altering their values, while keeping all the other features unchanged, affects the prediction of a selected data point. 
We generate realistic, feasible and actionable counterfactuals with a
state-of-the-art explainer called FACE and display them as
text~\citep{poyiadzi2020face}.
As a result, all of our explanations are perfectly faithful with respect to the underlying ML models; because the models themselves are transparent, the explanation validity can be easily verified. 

Given that the explanations differ in their scope and the type of information they contain, it is impractical to communicate them through a single modality while simultaneously ensuring that the information is presented clearly and concisely. Therefore, each explanation is represented with its most common modality as reported in existing literature. 
Feature importance is shown as a bar chart, as used by the popular
SHAP~\citep{lundberg2017unified} and LIME~\citep{ribeiro2016should} explainers.
We use a two-dimensional visualisation to show a decision surface, following
the common practice~\citep{huysmans2011empirical,goldstein2015peeking}.
Counterfactual explanations are predominantly offered in the form of a textual
statement~\citep{wachter2017counterfactual,russell2019efficient}. For the model
structure of a decision tree, we use its canonical, hierarchical
representation~\citep{lim2009and,bell2022s}. To display the structure of a
logistic regression model, we present both its mathematical formulation
(equations) as well as a table with its coefficients to assist users who lack
mathematics background, following similar practice reported in the XAI
literature~\citep{bell2022s}. 

Since these explanations cannot be delivered in the same modality,
we set the explanation technique as one of the independent variables in our study. 
We are not interested in finding the explanation modality that is the most understandable, but rather in investigating
whether different types of explanation techniques, which convey different information, are susceptible to over-interpretation. 
If we fix the modality across explanation types, their quality is compromised and their intelligibility degraded, which 
risks creating unintelligible explanations for the sake of controlling their modality. 
Additionally, using a modality of an explanation that differs from the one used in real life is likely to impair the generalisability and ecological validity of our study. 
We therefore present each explanation in its most common and intelligible modality.

\subsection{Mixed Factorial Design}\label{sec:factors}

Following the discussion in Section~\ref{sec:ml-model} and Section~\ref{sec:explanations}, we use a mixed factorial design in our experiment, with two ML models -- logistic regression or decision tree -- as between-subjects factors, and four explanation types -- model architecture, decision surface visualisation, feature importance and counterfactual explainability -- as within-subjects factors. 
By varying the ML models, we aim to explore if the model type plays a role in explanation comprehensibility.
Among the four explanations for the same ML model, we use the model architecture as the baseline to study the extent to which the model-independent explanations improve or deteriorate comprehensibility compared with presenting the raw model itself.

\subsection{Use Case}%

We place our user study in the context of chronic disease diagnosis, where
data-driven predictive models have high impact and are subject to
regulation~\citep{aida2023}, thus requiring validation and
certification~\citep{voigt2017eu}. 
Predictive models deployed in clinical diagnostics are further subject to additional
regulatory scrutiny such as the European In Vitro Diagnostic Medical Devices
Regulation, which emphasises the need for XAI tools in the medical
domain~\citep{muller2022explainability}.
Similar framing has also been used in previous XAI
research~\citep{binns2018s,schoonderwoerd2021human}. 
Additionally, using inherently interpretable models is desirable in high stakes
domains such as
healthcare~\citep{rudin2019stop,ghassemi2021false,jiang2022needs}. 
Evaluating the intelligibility of explanations in this context is therefore informative and ensures the ecological validity of our findings. 

With the emergence of digital healthcare systems, it is important to make such systems informative and accessible to their lay users. 
Our use case adheres to a common scenario where an at-home healthcare system relies on easy-to-collect biomarkers to help a user keep track of their health status. 
Specifically, a data-driven model informs a person if they are at a high risk of developing a chronic disease, which is intended to prompt them to seek professional medical advice. 
By explaining such algorithmic predictions, users do not need to understand the physiological processes that underlie a particular medical condition but rather know how to manage it. 
Therefore, data features should be familiar to non-experts and their number manageable. 
The target audience of our study is thus the lay population (not medical professionals) who do not necessarily have knowledge of medicine or AI. 

To support the external validity and reproducibility of the study, we use the
UCI diabetes dataset~\citep{smith1988using}, which is a real-world dataset
frequently used in machine learning
research~\citep{choubey2017classification,kumar2017performance}. The target
variable in the original dataset is to predict whether a patient has diabetes
based on eight diagnostic measurements such as the number of pregnancies, BMI,
insulin level and age.  
To generalise the context of our study, we remove the \emph{number of
pregnancies} feature -- since it is a gender-specific variable -- leaving us a
total of seven features. This feature set size is compatible with the human
cognitive capacity as according to \citet{miller1956magical}, humans can
simultaneously hold about seven
items in working memory.

In our experiments, we also modify the target variable in the binary classification task from \emph{having diabetes} (label~1) or not (label~0) to \emph{having a high risk} (label~1) or low risk (label~0) of developing the disease as we believe that the latter task is closer to a real-world use case of data-driven tools in healthcare.
We further reuse the models trained on the diabetes dataset across a series of scenarios where we introduce them as being deployed for the diagnosis of different chronic diseases, and specifically, to predict whether a person has a high or low risk of developing such a condition. 
This set-up allows us to generalise the scope of our user study and avoid the priming effect (see  Section~\ref{sec:priming} for a detailed discussion). 

The logistic regression model uses all seven features and achieves 81\% accuracy; the decision tree model achieves the highest accuracy of 77\% for which it only requires three of the features.
We generate each explanation for a data point sourced from the dataset, making sure that it is classified correctly by the underlying model, which ensures the validity of the explanatory insights. 

\subsection{Pilot Study}

We conducted two rounds of pilot studies with 12 researchers from our research centre to fine-tune the explanations and the questionnaire used to evaluate user comprehension with the help of the participants' feedback.  
We did this to validate the participants' understanding of the survey procedure 
and to refine the wording of our questions, which helped us to 
control the difficulty of each question and eliminate any confusion. 
We also assessed their perception of the survey workload to determine the appropriate number of questions to ensure that sufficient data can be collected without the risk of user fatigue.  

During the initial design phase of our experiment, we included two distinct scenarios, both in the healthcare domain but of different impact levels: a high stakes scenario focused on chronic disease diagnosis and a lower stakes scenario concerned with a general health check. We attempted to evaluate whether user perception of explanations would diverge when the impact of the model differed while controlling for the confounding factor of the application domain. 
Most of the pilot study participants pointed out that they perceive both scenarios as high stakes despite the chronic disease diagnosis having bigger ramifications than advice to undergo a general health check. 
This feedback prompted us to drop the lower stakes use case in the final study. 
Finally, we ran two additional rounds of the pilot study on the crowd-sourcing platform Prolific with a total of 15 participants to validate the survey procedure and collect feedback from our target demographics.

\begin{figure}[t!]
    \centering
    \includegraphics[width=\linewidth]{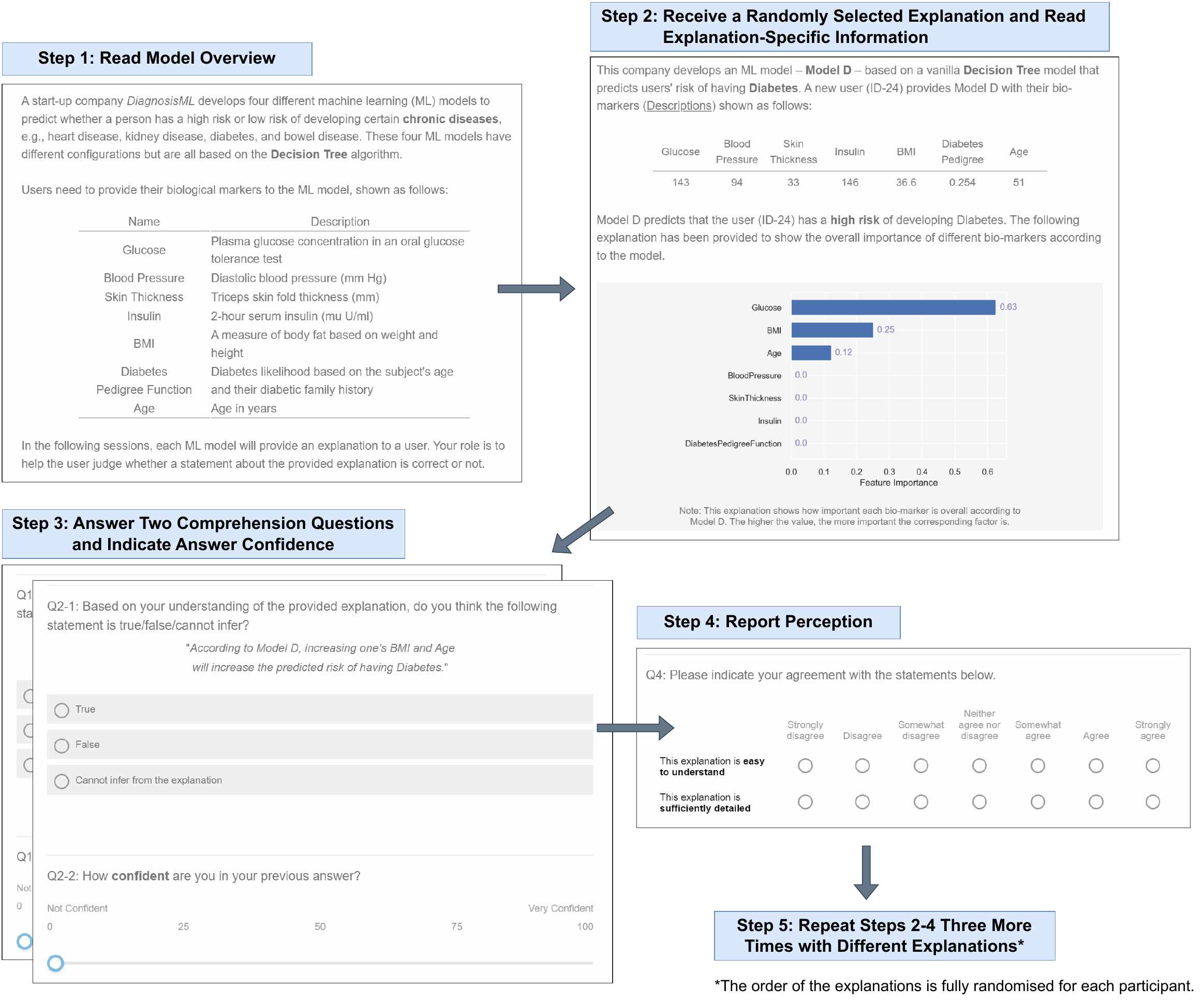}
    \caption{
    Overview of the user study workflow employed to assess user comprehension and perception of different explanations. In the actual survey, Step~1 is shown in a separate screen; once a participant clicks \emph{Next}, Steps~2--4 are shown in sequence in a single new screen. Clicking \emph{Next} again leads to an updated screen showing Steps~2--4 for a new explanation. For each explanation, the participant can hover their mouse over the \underline{Descriptions} keyword (visible in Step~2), which triggers a drop-down box showing information from Step~1. Therefore, the participant can always revisit model information in case they forget the details.
    \label{fig:main-procedure}
    }
\end{figure}

\section{User Study}\label{sec:user-study}

As discussed in Section~\ref{sec:factors}, our user study has eight independent
variables. Given that each variable requires a sample size of $n=20$ to
achieve high statistical power~\citep{van2014modern}, and to be conservative,
we decided to recruit 200 participants. 
Before reporting our findings, we describe our participants, the study procedure and the design of our comprehension questions. The study has been approved by our institutional Human Ethics Advisory Committee.

\subsection{Participants}

Participants were recruited through the Prolific platform. To ensure sufficient
response quality, we restricted participation to subjects with an acceptance
rate of 95\% or above, following the procedure outlined
by~\citet{van2021effect}. We also limited the study to crowd workers based in
the United Kingdom (UK). We used quotas for gender to ensure an even
distribution of male and female participants. After accepting the task, the
participants were directed to the Qualtrics platform, which hosted our
experiment. 
Adhering to the UK minimum wage of \pounds{}10.42 an hour at the time of our study and an expected completion time of 15~min (determined based on our pilot data), we compensated each participant with \pounds{}2.75 (equivalent to \pounds{}11 per hour) for completing our study.
We included three attention-check questions in the experiment; 
the answers to these questions were presented on the same page so that no memory recall was required.
An example of such a question is ``Which information is provided by this user to the ML model?'', with possible responses ``(A) Insulin'', ``(B) Air Pollution'', ``(C) Smoking'' and ``(D) Vitamin Supplements''. We did not use responses from participants who failed two or more -- out of three -- attention-check questions, but they were still compensated for the task.

\subsection{Procedure}\label{sec:procedure}

The participants first saw the consent form that described the study and the tasks involved. After consenting, the participants were assigned to one of two conditions at random: logistic regression or decision tree.
We restricted participants from taking part in our experiment more than once, yielding 101 participants in the logistic regression and 99 in the decision tree condition. 
A sample questionnaire received by our participants is shown in Figure~\ref{fig:main-procedure}.\footnote{The survey data and code used for our analysis are available at
\url{https://github.com/xuanxuanxuan-git/hcxai}.}
The high-level survey flow can be found in Figure~\ref{fig:flow-chart-simple} in Appendix~\ref{app:flow}, with the main steps outlined below.

\begin{enumerate}
    \item \textbf{Model overview.} The participants first read an introduction and a high-level summary of the predictive task -- chronic disease diagnosis -- along with information about four different configurations of an ML model -- either logistic regression or decision tree -- that were developed; each configuration corresponds to a different disease diagnosis to mitigate the priming effect (see Section~\ref{sec:priming} for a detailed discussion). 
    We also provided a description of the data features (see Step~1 in Figure~\ref{fig:main-procedure}). The participants were then informed about the tasks they needed to perform.
    \item \textbf{Evaluation questionnaires.}  \label{item:evalq}
    We evaluate each explanation -- specifically, its intelligibility -- from two perspectives: users' objective understanding and subjective assessment, which is popular in the XAI literature~\citep{cheng2019explaining,bove2022contextualization,bove2023investigating}. 
    \begin{itemize}
    \item \textbf{Read explanation-specific information.} 
    The participants were presented with one of the four ML explanation types selected at random -- model architecture, decision surface visualisation, feature importance or counterfactual explainability -- and a short description of the purpose of this explanation (Step~2 in Figure~\ref{fig:main-procedure}). Each explanation was generated to explain a model output corresponding to a random data profile, i.e.,~a data point, such that each explanation delivered different information. More details about how each explanation is introduced are provided in Section~\ref{sec:exp-disp}. The participants were then asked to complete the following questionnaire.
    \item \textbf{Comprehension questions.} 
        We employed comprehension questions (described below) to assess participants' objective comprehension and misinterpretation of ML explanations. 
        Each comprehension question consisted of an assertion that the participants were asked to judge based on their understanding of an explanation (Step~3 in Figure~\ref{fig:main-procedure}). Specifically, we asked: ``Based on your understanding of the provided explanation, do you think the following statement is true/false/cannot infer?'', with the possible answers being ``true'', ``false'' and ``cannot infer from the explanation''. 
        As discussed in Section~\ref{sec:explanations}, the information explicated by an explanation was determined by its scope and type, therefore each explanation required bespoke comprehension questions. 
        We designed two types of questions -- comprehension of \emph{explicated} and \emph{unspecified} information -- to capture two aspects of user understanding -- explanation comprehension and over-interpretation. 
        \begin{itemize}%
            \item \textbf{Comprehension of \emph{explicated} information.} We measured whether participants can understand the insights communicated by an explanation through assertions about \emph{explicated} information. 
            The factual correctness of these assertions can be determined solely based on the information provided by the corresponding explanations. In other words, the answers to assertions about \emph{explicated} information can only be ``true'' or ``false''.             
            Examples of such assertions are:
            \begin{itemize}
                \item ``model~D uses all 7 biomarkers while making predictions, and Glucose is the most important factor for model~D'' for \emph{feature importance}; and
                \item ``assuming that all other biomarker values remain the same (including BMI), increasing this person's Glucose to 135 will change their prediction from low risk to high risk'' for \emph{decision surface visualisation}.
            \end{itemize}
            For these examples, the former statement was ``true'' while the latter was ``false''; 
            participants were assigned to either a ``true'' or a ``false'' statement at random for this type of question.
            \item \textbf{Comprehension of \emph{unspecified} information.} We also measured whether participants know what information an explanation does not communicate to assess whether they can recognise the limitations of explanatory insights. To this end, we designed assertions that, while relevant to the explanations, could not be answered based on their content since information needed to judge these assertions was unspecified by the explanations. Examples of such assertions are: 
            \begin{itemize}
                \item ``BMI and Glucose are the MOST influential factors (among all 7 factors) in determining this person's result'' for \emph{decision surface visualisation}; and
                \item ``according to Model~D, increasing one's BMI and Insulin would increase the predicted risk of having Diabetes'' for \emph{feature importance}.
            \end{itemize}
             The correct answer to all questions in this category is ``cannot infer from the explanation''.
        \end{itemize}%
        In summary, for each explanation, participants were shown one question about explicated information (to which the correct answer was either ``true'' or ``false'') and one question about unspecified information (to which the correct answer was ``cannot infer''). The order in which the two questions were displayed was randomised. The complete list of comprehension questions for each explanation is provided in Appendix~\ref{app:exp}. The design principles of our comprehension statements are discussed in  Section~\ref{sec:statement}.     
    \item \textbf{Answer confidence.} Every quantitative (objective) comprehension question was accompanied by a request to report the participant's confidence in their answer on a range of 0 to 100 (Step~3 in Figure~\ref{fig:main-procedure}), following a similar practice in other XAI user studies~\citep{huysmans2011empirical}. 
    \item \textbf{Explanation perception.} We asked participants to report their agreement with ``this explanation is \emph{sufficiently detailed}'' on a 7-point Likert scale, ranging from ``strongly disagree (1)'' to ``strongly agree (7)'', which question was adapted from a similar user study~\citep{hoffman2018metrics}. Participants were also asked to report their perceived difficulty in comprehending the explanation by rating their agreement with ``the explanation is \emph{easy to understand}'' using the same scale, which question was adapted from a similar user study~\citep{lage2019human}. See Step~4 in Figure~\ref{fig:main-procedure} for reference. 
    \item \textbf{Repeat.} Participants repeated all the actions listed under the \emph{Evaluation questionnaires} item with a different and randomly selected explanation for a total of four times per experiment.  
    Each explanation was introduced in a distinct medical context -- diagnosis of a different disease -- for a different individual -- previously unseen data instance -- to prevent participants from accumulating information about the underlying predictive model after interacting with previous explanations.
    \end{itemize}
    \item \textbf{Graph literacy.} We assessed the graph literacy of the participants using the Short Graph Literacy (SGL) scale~\citep{okan2019using}, following the set-up reported in prior XAI research~\citep{van2021effect}. SGL is a validated questionnaire consisting of four questions, each based on a visual graph.
    \item \textbf{Demographic information.} We asked participants to report their age, gender and educational attainment. In addition to the basic demographic information, we also asked participants to report their literacy in machine learning algorithms, English proficiency level, technical background and their experience with similar XAI user studies. To evaluate \emph{algorithm literacy}, we asked participants to report their knowledge of ML algorithms on a 5-point Likert scale; we adapted this question from a prior study~\citep{wang2020factors}. To measure users' \emph{technical background}, we asked participants to report their participation in the field of Science, Technology, Engineering and Mathematics (STEM); we adapted this question from a prior study as well~\citep{reeder2023evaluating}. We also asked users to report their \emph{English proficiency level}, which was categorised into six proficiency levels according to the Common European Framework of Reference for Languages (CEFR)~\citep{persons2001common}. 
    Lastly, we asked participants to report their previous participation in similar XAI user studies, ranging from ``None (0)'' to ``A lot (probably more than 15)''.
\end{enumerate}

\subsection{Explanation Display}\label{sec:exp-disp}
Given that each explanation is generated for a specific ML predictor and an individual prediction result, we always provided the context and usage of an explanation before introducing it to the participants. For each explanation, we first described the predictor and its usage (e.g.,~predicting the risk of having diabetes). Then, we provided the data profile of an anonymous individual and the prediction result they received from the ML model (e.g.,~high risk or low risk of having diabetes). Next, we provided an explanation along with a short textual description of the explanation in a grey box for improved visualisation; for example, feature importance was described as: ``This explanation shows how important each biomarker is overall according to Model~D. The higher the value, the more important the corresponding factor is.'' Within the same web page, the participants were asked to complete the questions for objective and subjective assessment. Clicking the \emph{Next} button directed the participants to the next explanation, which dealt with a new ML predictor and data profile. The participants could not go back to see previous explanations as these are independent of each other and the information they provide should be analysed separately.
All the explanations used in our study and their description are shown in Appendix~\ref{app:exp}. An example of a complete questionnaire distributed to a participant is shown in Figure\ref{fig:flow-chart-example} given in Appendix~\ref{app:flow}.

\begin{table}[t!]
    \centering
\small
    \caption{Overview of explicated and unspecified information for each explanation assessed by our comprehension assertions.}
    \begin{tabular}{@{}p{1.4cm}p{2.5cm}p{2.5cm}p{2.5cm}p{2.5cm}@{}}
    \toprule
         & Feature importance & Decision surface & Counterfactual & Model architecture \\ \midrule %
    Scope & Global & Local & Local & Global \\
      Explicated information & Global feature importance & Change in the prediction based on values of \emph{two} features & Effect of a feature vector change on the prediction & Model structure (simulatability) \\
      Unspecified information & Local feature influence & (Local) importance of the two displayed features & (Local) importance of the two listed features & (Combined) impact of features \\
      \bottomrule
    \end{tabular}
    \label{tab:question_design}
\end{table}

\subsection{Comprehension Questions}\label{sec:statement}%

The statements used in the questionnaire to evaluate objective comprehension were distinct for each explanation type. These questions quizzed participants about their core understanding of both the information explicated and unspecified by an explanation. 
Specifically, the statements about \emph{unspecified information} asked participants about details that were relevant to an explanation but could not be uniquely determined. 
For example, our decision surface visualisation communicates how changes in the values of two features influence the prediction of the underlying model for a particular user, but such an explanation lacks information about the importance of these two features. For this explanation, we employed a question to assess whether participants can predict the new output of the model given different values of these two features as \emph{explicated information}; and another question about whether they incorrectly inferred \emph{unspecified information} -- feature importance -- from this specific explanation. 
The rationale behind the latter question is that unspecified information can be easily misconstrued in this case. 
Our design -- informed by an XAI expert and improved iteratively during the pilot studies -- ensures that the questions resemble each other and are comparably difficult to judge across explanation types.

Given that local explanations -- such as decision surface visualisation -- provide insights that are necessarily limited in scope, a wide range of information remains (implicitly) unspecified by this explanation type. Since all four explanation types used in our study entailed insights about features, we designed comprehension questions purely around this explanatory aspect. 
Specifically, the statements either quizzed the participants about the output of the model given a particular change to the feature vector or about feature importance and influence. An overview of explicated and unspecified information relevant to each explanation type is provided in 
Table~\ref{tab:question_design}.
While focusing predominantly on feature-based explanatory information implicitly limits the scope of our results, our study offers a principled evaluation of explicated and unspecified explanatory information that is otherwise overlooked by the current literature.

\subsection{Priming Effect}\label{sec:priming}

Given that the explanation type is a within-subjects factor, it is crucial to address the priming effect in our study design. To this end, we fully randomised the order of four explanation types presented to each participant. To remove the risk of participants accumulating information across tasks, 
we framed each explanation in the context of a different disease diagnosis. Specifically, for each participant, the ML model was presented in four different configurations, each under a different pseudo-name such as ``Model~B'' and 
tasked with diagnosing the risk of one of the four diseases: diabetes, heart disease, kidney disease and bowel disease. 
We further introduced a new user profile, with its unique biomarkers, for which an explanation was generated for each explanation type. Each such profile was assigned with a different user ID, e.g.,~ID-42. The use case description also emphasised that each explanation was provided in a different context. By doing so, participants had limited chances of learning between explanations. After collecting the survey responses, we further checked the priming effect and verified that our participants' performance on the comprehension questions did not increase as they progressed through the survey (see Appendix~\ref{app:priming} for more details). 

\begin{table}[t!]
\centering
\small
\caption{Overview of major personal characteristics for our 200 participants.}
\begin{tabular}{@{}wc{4.0cm}p{5cm}rr@{}}
\toprule
Characteristic & Values
& Frequency & Percentage \\ \midrule %
\multirow{3}{*}{Gender} & Female & 99 & 49.5\% \\
 & Male & 96 & 48.0\% \\
 & Non-binary & 5 & 2.5\% \\\cmidrule(lr){1-4}%
\multirow{6}{*}{Age} & 18--24 years old &  37 & 18.5\% \\
 & 25--34 years old & 49 & 24.5\% \\
 & 35--44 years old & 44 & 22.0\% \\
 & 45--54 years old & 31 & 15.5\% \\
 & 55--64 years old & 22 & 11.0\% \\
 & 65+ years old   & 17 & 8.5\% \\ \cmidrule(lr){1-4}%
\multirow{5}{*}{Educational attainment} & Less than high school degree & 4 & 2.0\%\\
 & High school graduate & 38 & 19.0\% \\
 & College degree & 30 & 15.0\% \\
 & Bachelor's degree & 86 & 43.0\% \\
 & Graduate or professional degree & 42 & 21.0\% \\ \cmidrule(lr){1-4}%
\multirow{4}{*}{ML algorithm literacy} & No knowledge & 95 & 47.5\% \\
 & Negligible knowledge & 63 & 31.5\% \\
 & Some knowledge & 35 & 17.5\% \\
 & Moderate knowledge & 7 & 3.5\% \\ \cmidrule(lr){1-4}%
\multirow{2}{*}{Technical background} & Yes (STEM-related education/employment) & 61 & 30.5\% \\ %
& No & 139 & 69.5\% \\  \cmidrule(lr){1-4}%
\multirow{5}{*}{Experience with XAI user study} & None (0) & 125 & 62.5\% \\
 & A few (roughly 1--5) & 69  & 34.5\% \\
 & A fair amount (around 6--15) & 4 & 2.0\% \\
 & A lot (probably more than 15) & 2  & 1.0\% \\
\bottomrule
\end{tabular}
\label{tab:demographic}
\end{table}

\section{Results}\label{sec:results}

201 participants completed the study from which 200 responses were included in the final analysis because one participant failed the attention check. 
The demographic information of our participants is summarised in Table~\ref{tab:demographic}. 
The average completion time of our study was 13.35~mins, with a minimum
completion time of 4.72~mins. The participants' average graph literacy score --
assessed through the Short Graph Literacy scale -- was 2.46 ($\mathit{SD}$ $=$ 1.06,
ranging from 0 to 4), which is close to the average score of 2.2 observed by
the original study done on the US population~\citep{okan2019using}. 79\% of the
participants indicated that they have ``no knowledge'' or ``negligible
knowledge'' of machine learning algorithms; 69.5\% of the participants
indicated that they do not have STEM background. 
The population of our respondents is aligned with the envisaged target audience, namely stakeholders who lack technical expertise.

\begin{table}[t]
\centering
\small
\caption{%
Coefficients, standard errors (in brackets) and significance indicators ($^{\star{}}$ for $p< 0.05$, $^{\star{}\star{}}$ for $p<0.01$ and $^{\star{}\star{}\star{}}$ for $p<0.001$) of
statistical regression models used to assess the effect of ML model type and explainability approach on user comprehension. 
Models~1 and 3 show main effects on user comprehension of explicated and unspecified information respectively; Models~2 and 4 include interaction terms.
Note that these statistical regression models are used to explore the correlation between variables (i.e.,~model type and explainability approach) and user comprehension, not for predicting user comprehension, therefore we focus more on their coefficients and $p$-values rather than the goodness of their fit (Pseudo-$R^2$).}
\begin{tabular}{rllll}
\toprule
 & \multicolumn{2}{c}{Explicated information} & \multicolumn{2}{c}{Unspecified information}\\ \cmidrule(lr){2-3}\cmidrule(lr){4-5} 
 & \multicolumn{1}{c}{Model 1} & \multicolumn{1}{c}{Model 2} & Model 3 & Model 4 \\
 \midrule
Model type: Logistic regression & 0.18 (0.15) & 0.04 (0.28) & 0.32 (0.16)$^{\star}$ & 0.59 (0.33)\\
Explanation type: Feature importance & 0.74 (0.21)$^{\star\star\star}$ & 0.50 (0.29) & -0.39 (0.24) \; & -0.34 (0.37)\\
Explanation type: Decision surface & 1.75 (0.24)$^{\star\star\star}$ & 1.75 (0.35)$^{\star\star\star}$ & -0.16 (0.23) \; & 0.00 (0.35) \\
Explanation type: Counterfactual  & 0.10 (0.20) & 0.08 (0.29) & 0.53 (0.22)$^{\star}$ & 0.84 (0.32)$^{\star\star}$\\
Logistic regression + Feature importance & \multicolumn{1}{c}{---} & 0.50 (0.42) & \multicolumn{1}{c}{---} & -0.11 (0.49)\\
Logistic regression + Decision surface & \multicolumn{1}{c}{---} & -0.02 (0.49) & \multicolumn{1}{c}{---} & -0.29 (0.47)\\
Logistic regression + Counterfactual  & \multicolumn{1}{c}{---} & 0.04 (0.40) & \multicolumn{1}{c}{---} & -0.58 (0.44)\\
\midrule
Pseudo-$R^2$ & \multicolumn{1}{c}{0.07} & \multicolumn{1}{c}{0.08} & \multicolumn{1}{c}{0.03} & \multicolumn{1}{c}{0.03} \\
\bottomrule
\end{tabular}\label{tab:log_stats_model}
\end{table}

\begin{figure}[t]
    \centering
    \begin{subfigure}[b]{0.33\textwidth}
        \centering
        \includegraphics[width=\textwidth]{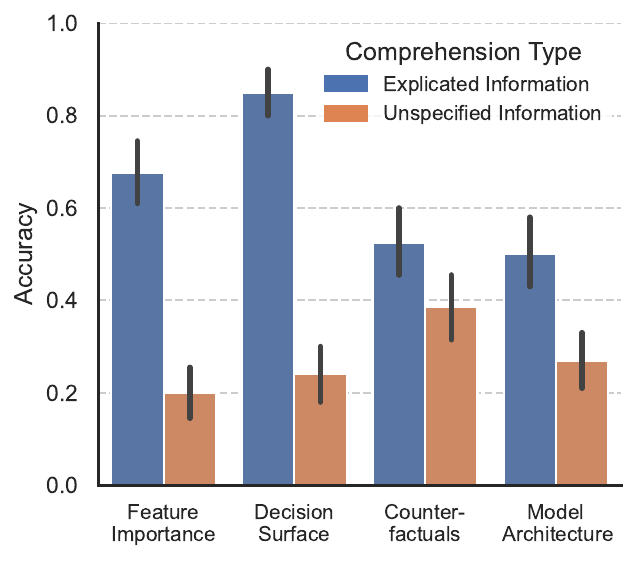}
        \caption{Explanation type.}\label{fig:overall_compare_0}%
    \end{subfigure}
    \hspace{.15em}
    \begin{subfigure}[b]{.185\textwidth}
    \centering
     \includegraphics[width=\textwidth]{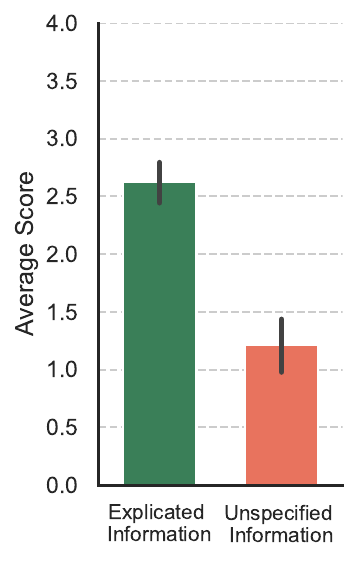}
    \caption{Logistic regression.}\label{fig:overall_compare_1}%
    \end{subfigure}
    \hspace{0.15em}
    \begin{subfigure}[b]{.165\textwidth}
    \centering
     \includegraphics[width=\textwidth]{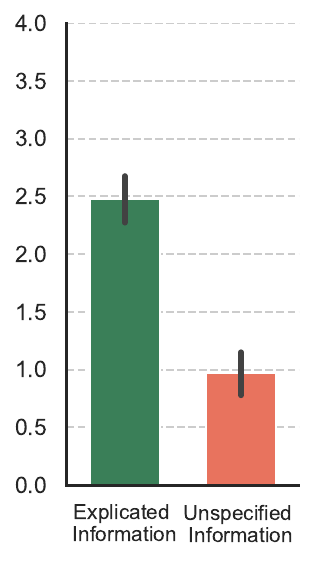}
    \caption{Decision tree.}\label{fig:overall_compare_2}%
    \end{subfigure}
    \hspace{0.15em}
    \begin{subfigure}[b]{.14\textwidth}
    \centering
     \includegraphics[width=\textwidth]{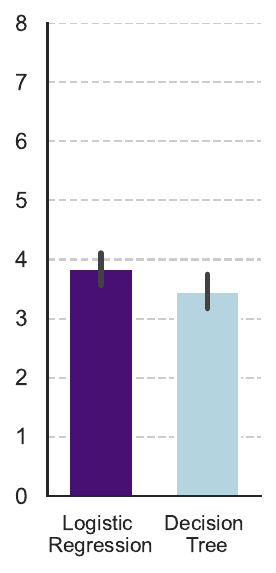}
    \caption{Overall score.}\label{fig:overall_compare_3}%
    \end{subfigure}
    \caption{Accuracy of the participants' answers to the question about explicated and unspecified information stratified by explanation type is shown in Panel~(\subref{fig:overall_compare_0}). Participants were significantly more likely to understand explicated information shown in feature importance and decision surface visualisation but less attuned to their unspecified information, in contrast to counterfactuals and model architecture. The average user comprehension score for the four questions about explicated information and the four questions about unspecified information grouped by explanation type is shown in Panel~(\subref{fig:overall_compare_1}) for logistic regression and Panel~(\subref{fig:overall_compare_2}) for decision tree. Participants were significantly more likely to have correct comprehension of the information unspecified by the explanations of logistic regression compared to decision tree. Average score for all comprehension questions (including all eight questions about explicated and unspecified information) stratified by the ML model type is shown in Panel~(\subref{fig:overall_compare_3}). All error bars indicate 95\% confidence interval.}
    \label{fig:overall_compare}
\end{figure}

\begin{table}[t]
    \centering
\small
    \caption{$\mathcal{X}^2$ and significance indicators ($^{\star{}}$ for $p< 0.05$, $^{\star{}\star{}}$ for $p<0.01$ and $^{\star{}\star{}\star{}}$ for $p<0.001$) of McNemar's test used to assess whether user comprehension of explicated and unspecified information is identical across explanation types.}
    \begin{tabular}{@{}rllll@{}}
    \toprule
    & \multicolumn{4}{c}{Explanation Type} \\ \cmidrule(lr){2-5}
    & Feature importance & Decision surface & Counterfactual & Model architecture \\ %
    \midrule
    Logistic regression & 39.06$^{\star\star\star}$ & 49.01$^{\star\star\star}$ & 3.95 $^{\star}$ & 4.50$^{\star}$ \\
    Decision tree  & 35.53$^{\star\star\star}$ & 57.52$^{\star\star\star}$ & 3.07 & 17.82$^{\star\star\star}$  \\ 
    \midrule
    All & 74.59$^{\star\star\star}$ & 106.31$^{\star\star\star}$ & 7.0$^{\star\star}$ & 18.24$^{\star\star\star}$ \\ \bottomrule
    \end{tabular}
    \label{tab:mc_tests}
\end{table}

\begin{figure}[t]
    \centering
    \begin{subfigure}[t]{0.30\linewidth}
    \centering
     \includegraphics[height=\textwidth]{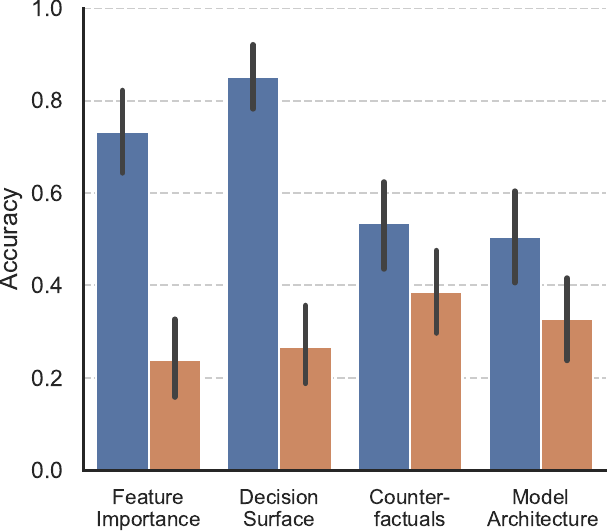}%
    \caption{Logistic regression.}\label{fig:comp_two_models_0}
    \end{subfigure}
    \hspace{5em}
    \begin{subfigure}[t]{0.3\linewidth}
    \centering
     \includegraphics[height=\textwidth]{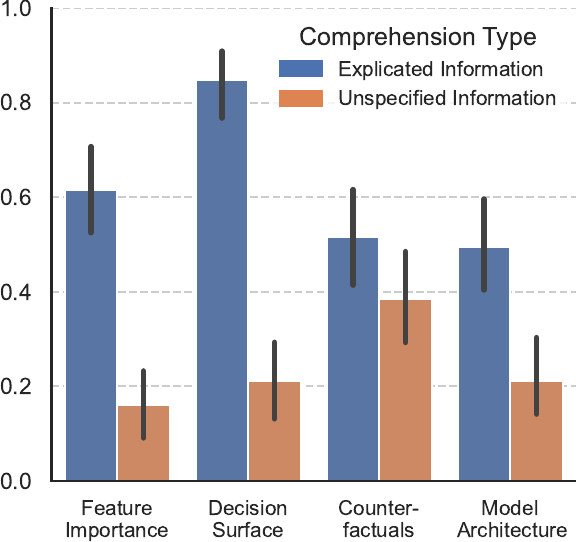}%
    \caption{Decision tree.}\label{fig:comp_two_models_1}
    \end{subfigure}
    \caption{Accuracy of our participants' answers to comprehension questions about explicated and unspecified information grouped by explanation type, separately for our two ML models. Error bars indicate 95\% confidence interval. The participants' comprehension of explicated information is substantially more accurate than their comprehension of unspecified information for every explanation type of logistic regression and three explanation types of decision tree (except counterfactual explainability).}
    \label{fig:comp_two_models}
\end{figure}

\subsection{User Comprehension}

In order to identify the effect of \emph{ML model type} and \emph{explanation type} on the participants' comprehension of explicated and unspecified information, we constructed statistical regression models to analyse the correlations. 
In each statistical model, the ML model and explanation type are the independent variables, and the dependent variable is whether a participant's answer to the comprehension question about explicated or unspecified information is correct or not. The modelling outcomes are summarised in Table~\ref{tab:log_stats_model}. 
Models~1 and 3 examine the main effects of ML model and explanation type on
user comprehension of explicated and unspecified information respectively.
Models~2 and 4 include the interaction effects between ML model and explanation
type. To ensure the validity of statistical modelling, we checked for the
existence of multicollinearity among model parameters. We found that the
variance inflation factors of the parameters are in the 1--1.5 range, which is
below the threshold of 5 or 10 indicative of
multicollinearity~\citep{hair2009multivariate}. 

Model~1 in Table~\ref{tab:log_stats_model} shows that feature importance and decision surface visualisation have significant main effects on the user comprehension of \emph{explicated information}, which is visualised in  Figure~\ref{fig:overall_compare_0}. 
Additional McNemar's tests between pairs of explanations confirm that the participants were significantly more likely to have accurate comprehension of explicated information communicated by feature importance ($\mu$ $=$ 0.68, $\mathit{SD}$ $=$ 0.47) and decision surface visualisation ($\mu$ $=$ 0.85, $\mathit{SD}$ $=$ 0.36); this is in contrast to counterfactual explainability ($\mu$ $=$ 0.53, $\mathit{SD}$ $=$ 0.50) and model architecture ($\mu$ $=$ 0.50, $\mathit{SD}$ $=$ 0.50). Statistics of our pairwise McNemar's tests can be found in Table~\ref{tab:pairwise_comp} included in Appendix~\ref{app:stats_results}. 
Model~2 shows no significant interaction effects between explanation type and ML model on comprehension of explicated information.

Model~3 shows that the ML model type has a significant main effect on user comprehension of \emph{unspecified information}, which is visualised in Figures~\ref{fig:overall_compare_1} \& \ref{fig:overall_compare_2}. Our results indicate that the participants were significantly more likely to recognise the limitations of logistic regression explanations, i.e.,~their unspecified information.
This means that when the participants were working with ML explanations of logistic regression, they were significantly more likely to identify the constraints of these explanations (but less likely to do so for decision tree). 
We do not find a significant difference in comprehension of explicated information between the two ML models. The overall comprehension of explanations between these ML models was not significant either as shown in  Figure~\ref{fig:overall_compare_3}.
Model~3 also indicates that counterfactual explainability has a significant effect on comprehension of unspecified information -- see Figure~\ref{fig:overall_compare_0} for a visual representation. Additional McNemar's tests confirm a significantly higher comprehension level for counterfactuals ($\mu$ $=$ 0.39, $\mathit{SD}$ $=$ 0.49) compared to model architecture ($\mathcal{X}^2$ $=$ 5.82, $p$ $=$ 0.02), decision surface visualisation ($\mathcal{X}^2$ $=$ 16.49, $p$ $<$ 0.001) and feature importance ($\mathcal{X}^2$ $=$ 18.75, $p$ $<$ 0.001). 
No significant interaction effects are observed for comprehension of unspecified information according to Model~4.

In addition to statistical regression modelling, we also performed statistical tests to measure whether user comprehension of explicated and unspecified information is identical for each explanation separately. The statistics of McNemar's tests with continuity correction are summarised in Table~\ref{tab:mc_tests}. Test results show that for every explanation of logistic regression the proportion of our participants who correctly identified explicated information was significantly different from the proportion of our participants who successfully recognised unspecified information. 
This significant difference is visualised in Figure~\ref{fig:comp_two_models_0} and indicates that the participants' comprehension of explicated information is substantially more accurate than their comprehension of unspecified information for every explanation type of logistic regression. A similar phenomenon can be observed for the explanations of decision tree other than counterfactual explainability as demonstrated in  Figure~\ref{fig:comp_two_models_1}. After aggregating the participants' responses for the two ML models, McNemar's tests confirm that the proportion of our participants who have developed correct comprehension of explicated information was statistically different from the proportion who accurately identified unspecified information for every explanation type.

\begin{figure}[t]
    \centering
    \begin{subfigure}[b]{0.32\linewidth}
    \centering
     \includegraphics[height=.85\linewidth]{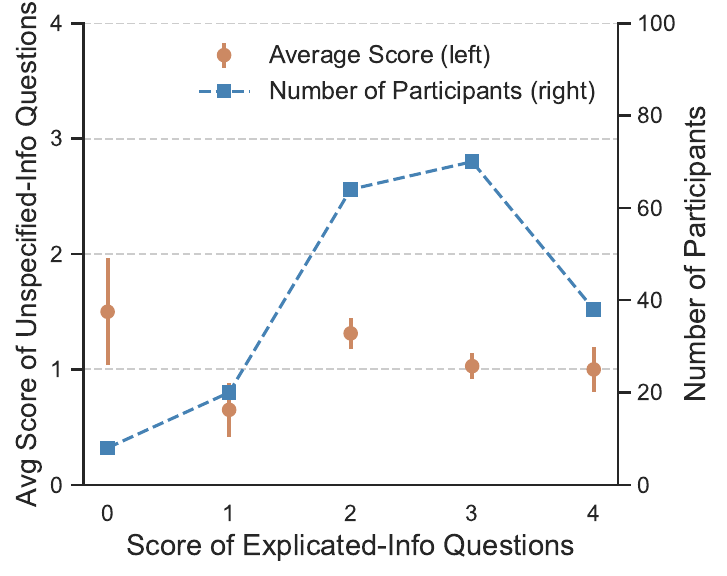}
    \caption{Performance comparison.
    }\label{fig:tf-ct-comp}
    \end{subfigure}
    \hfill
    \begin{subfigure}[b]{.32\linewidth}
    \centering
     \includegraphics[height=.85\linewidth]{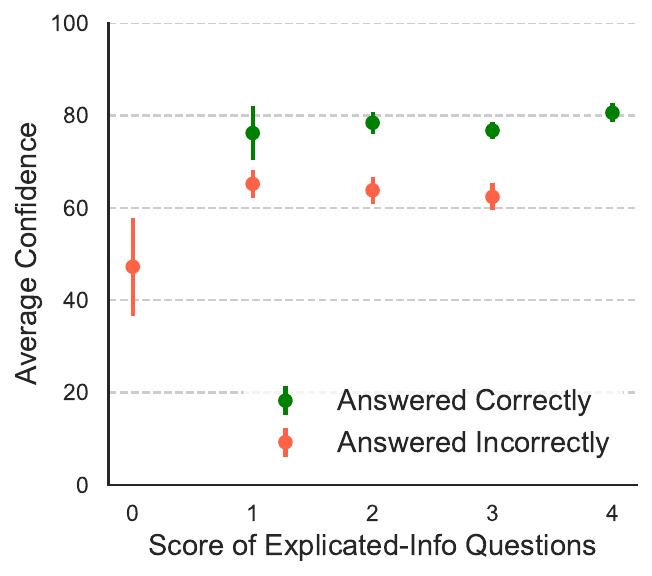}
    \caption{Explicated information.
    }\label{fig:tf-calibrated}
    \end{subfigure}
    \hfill
    \begin{subfigure}[b]{.32\linewidth}{
    \centering
     \includegraphics[height=.85\linewidth]{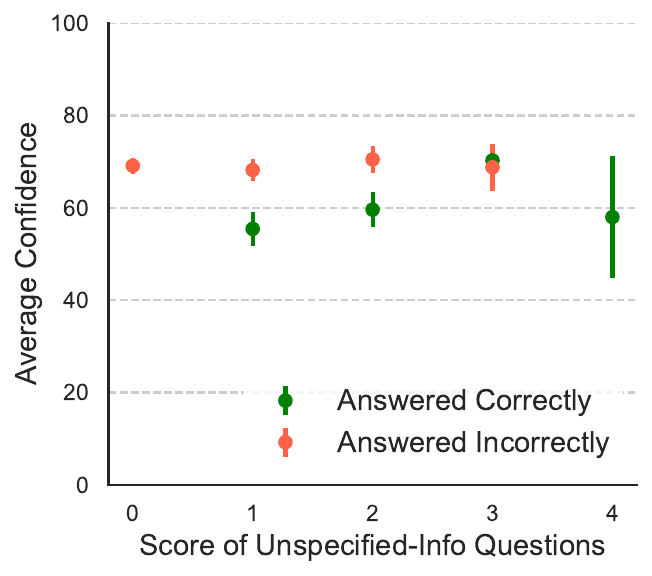}
    \caption{Unspecified information.
    }\label{fig:ct-calibrated}}
    \end{subfigure}
    \caption{
    Overview of our participants' performance for questions about explicated and unspecified information and their confidence level. Panel~(\subref{fig:tf-ct-comp}) suggests that the participants with different levels of performance on the questions about explicated information attained comparable average score for the questions about unspecified information. Panel~(\subref{fig:tf-calibrated}) shows participants with different levels of performance for questions about \emph{explicated information} and the confidence in their answers. Panel~(\subref{fig:ct-calibrated}) shows participants with different levels of performance for questions about \emph{unspecified information} and the confidence in their answers. Error bars indicate 95\% confidence interval.}
    \label{fig:tf-ct-comprehension}
\end{figure}

\begin{figure}[t]
    \centering
    \begin{subfigure}[b]{0.32\linewidth}
    \centering
     \includegraphics[width=\linewidth]{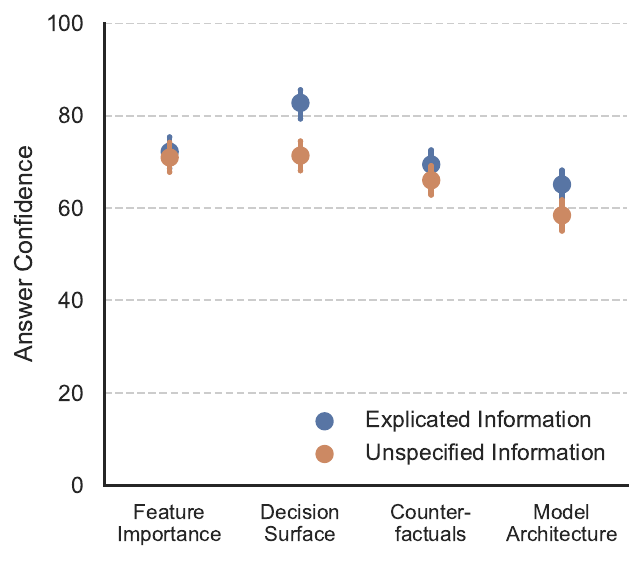}
    \caption{Aggregated.}\label{fig:confidence_0}%
    \end{subfigure}
    \hfill
    \begin{subfigure}[b]{.31\linewidth}
    \centering
     \includegraphics[width=\linewidth]{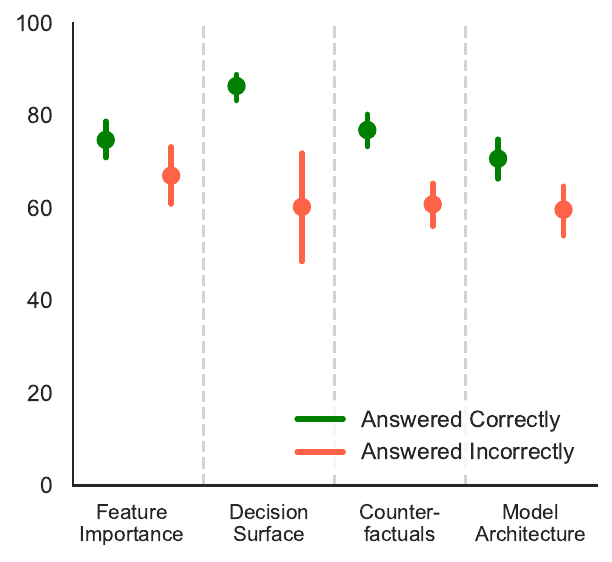}
    \caption{Explicated information.}\label{fig:confidence_1}%
    \end{subfigure}
    \hfill
    \begin{subfigure}[b]{.31\linewidth}
    \centering
     \includegraphics[width=\linewidth]{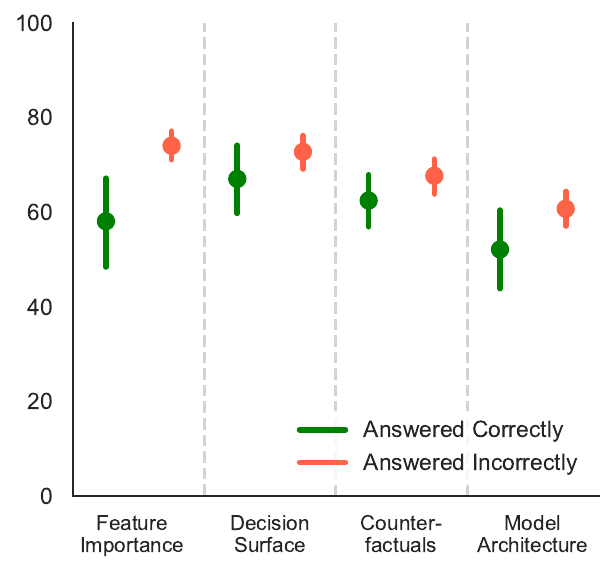}
    \caption{Unspecified information.}\label{fig:confidence_2}%
    \end{subfigure}
    \caption{
    Overview of our participants' confidence in their answers to comprehension questions about explicated and unspecified information is shown in Panel~(\subref{fig:confidence_0}). Answer confidence of participants who identified \emph{explicated information} correctly (displayed in green) and incorrectly (displayed in red) is shown in Panel~(\subref{fig:confidence_1}). Answer confidence of participants who identified \emph{unspecified information} correctly (displayed in green) and incorrectly (displayed in red) is shown in Panel~(\subref{fig:confidence_2}). Participants who answered the questions about \emph{explicated information} correctly reported significantly higher confidence in their answers than their peers who did not. On the other hand, participants who answered the questions about unspecified information correctly were less confident than those who answered them incorrectly. Error bars indicate 95\% confidence interval.}
    \label{fig:confidence}
\end{figure}

\subsection{Comprehension and Confidence}

We further compare the participants' performance on the comprehension questions about explicated and unspecified information and their confidence level. 
As shown in Figure~\ref{fig:tf-ct-comp}, regardless of how well the participants performed in the questions about explicated information, they had consistently worse performance in identifying unspecified information. One-way ANOVA test confirms that the difference in average performance on the questions about unspecified information was insignificant across participants with different comprehension levels of explicated information ($F_{(200)}$ $=$ 1.98, $p$ $=$ 0.10). 

The participants' confidence in their answers to the comprehension questions is visualised in Figure~\ref{fig:confidence_0}. A paired Student's t-test ($t$ $=$ 6.90, $p$ $<$ 0.001) shows that  
our participants were significantly more confident in their answers to the comprehension questions about explicated information ($\mu$ $=$ 72.36, $\mathit{SD}$ $=$ 17.24) in comparison to the questions pertaining to unspecified information ($\mu$ $=$ 66.67, $\mathit{SD}$ $=$ 17.04) regardless of the correctness of their comprehension. Additional paired t-tests confirm that the aforementioned observation is significant for decision surface visualisation ($t$ $=$ 6.95, $p$ $<$ 0.001), counterfactual explainability ($t$ $=$ 2.07, $p$ $=$ 0.04) and model architecture ($t$ $=$ 4.41, $p$ $<$ 0.001); however, we do not observe a significant difference in answer confidence for feature importance. 

We further compare answer confidence between participants who exhibited correct comprehension and those who did not, which difference is visualised in Figures~\ref{fig:confidence_1} \& \ref{fig:confidence_2}. 
We find that participants who answered the questions about \emph{explicated information} correctly reported significantly \emph{higher confidence} in their answers than their peers who did not. 
This is confirmed with additional independent t-tests for each explanation type: feature importance ($t$ $=$ 2.2, $p$ $=$ 0.03), decision surface visualisation ($t$ $=$ 6.12, $p$ $<$ 0.001), counterfactual explainability ($t$ $=$ 5.27, $p$ $<$ 0.001) and model architecture ($t$ $=$ 3.13, $p$ $=$ 0.002). 
On the other hand, participants who answered the questions about \emph{unspecified information} correctly were \emph{less confident} than those who answered them incorrectly. 
Independent t-tests show that this difference in confidence of unspecified information was significant for feature importance ($t$ $=$ -4.02, $p$ $<$ 0.001) and model architecture ($t$ $=$ -2.20, $p$ $=$ 0.03), but insignificant for decision surface visualisation and counterfactual explainability.

We also explore if participants with varying objective comprehension performance have different confidence levels in their answers to the questions about \emph{explicated} and \emph{unspecified information}.
As shown in Figures~\ref{fig:tf-calibrated} \& \ref{fig:ct-calibrated}, participants with different levels of comprehension had consistent levels of confidence across their answers to both question types. This indicates that a higher level of comprehension does not guarantee better confidence calibration. 

\begin{figure}[t]
    \centering
    \includegraphics[width=\linewidth]{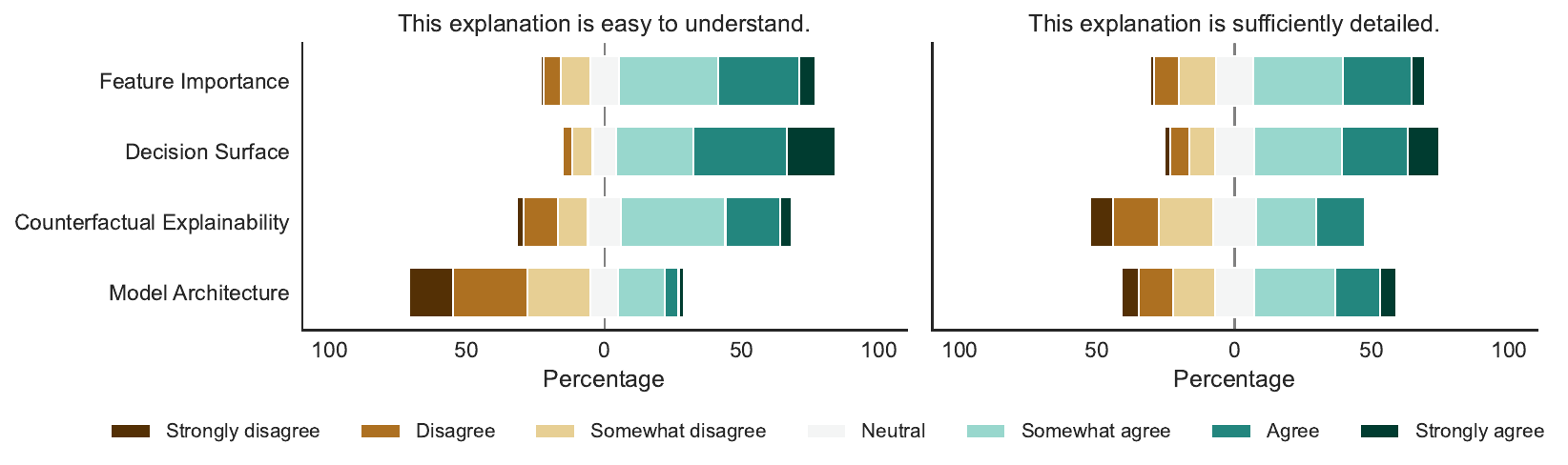}
    \caption{Response overview for the Likert scale questions stratified by explanation type. Across both questions, the perception was significantly more negative towards model architecture and counterfactual explainability than for feature importance and decision surface visualisation.}
    \label{fig:sub_comp}
\end{figure}

\subsection{Subjective Assessment}

Next, we analyse the participants' Likert-scale responses to ``this explanation is easy to understand'' and ``this explanation is sufficiently detailed'' stratified by explanation type, which are shown in Figure~\ref{fig:sub_comp}. 
A repeated measures ANOVA test shows that the user perception of comprehension
difficulty diverged significantly across explanation types
($F_{(3,591)}$ $=$ 121.61, $p$ $<$ 0.001).\footnote{While parametric tests such as
ANOVA are in principle only applicable to continuous data -- and Likert scale
offers discrete, ordinal data -- \citet{norman2010likert} showed that
parametric tests offer meaningful results that are generally more robust than
non-parametric tests in such situations, which practice has been adopted by
other XAI studies~\citep{binns2018s}. \citet{sullivan2013analyzing} also noted
that ordinal variables with five or more categories can often be used as
continuous without any negative impact on the analysis.} Specifically,
perception of model architecture and counterfactual explainability was
significantly more negative than of feature importance and decision surface
visualisation (see Tukey's post-hoc paired test results reported in Table~\ref{tab:easy_ttest} provided in Appendix~\ref{app:stats_results}). Additional repeated measures ANOVA tests also show that responses regarding the difficulty of understanding were significantly different across explanations if we evaluate them separately for logistic regression ($F_{(3, 300)}$ $=$ 101.0, $p$ $<$ 0.001) and decision tree ($F_{(3, 294)}$ $=$ 35.79, $p$ $<$ 0.001) -- see Figure~\ref{fig:easy_per_model} in Appendix~\ref{app:stats_results} for more details. 

In terms of the participants' appreciation of the explanations' richness of information, repeated measures ANOVA tests indicate that the responses diverged significantly across explanation types, both at an aggregate level ($F_{(3, 591)}$ $=$ 26.90, $p$ $<$ 0.001) as well as separately for logistic regression ($F_{(3, 300)}$ $=$ 13.54, $p$ $<$ 0.001) and decision tree ($F_{(3, 294)}$ $=$ 15.10, $p$ $<$ 0.001). Specifically, the participants were significantly more discontent with counterfactual explainability compared to the other explanation types -- see Table~\ref{tab:detail_ttest} and Figure~\ref{fig:detail_per_model} in Appendix~\ref{app:stats_results} for more details.

\begin{table}[t]
    \centering
    \small
    \caption{%
   Coefficients, standard errors (in brackets) and significance indicators ($^{\star{}}$ for $p< 0.05$, $^{\star{}\star{}}$ for $p<0.01$ and $^{\star{}\star{}\star{}}$ for $p<0.001$) of the 
    generalised linear models used to analyse the effect of individual characteristics on user comprehension scores. Models~5 and 7 show the main effect; Model~6 includes interaction terms.}
    \begin{tabular}{@{}rlll@{}}
    \toprule
         & Model 5   & Model 6          & Model 7 \\
         & (Aggregated) & (Explicated) & (Unspecified) \\
    \midrule
    Age &  0.00 (0.07) & 0.25 (0.18) & 0.01 (0.05) \\
    Gender: Male & 0.12 (0.21) & -0.22 (0.53) & 0.01 (0.16) \\
    STEM: Yes & 0.39 (0.24) & 2.31 (0.77)$^{\star\star}$ & 0.26 (0.18) \\
    Education & 0.33 (0.10)$^{\star\star}$ & 0.50 (0.21)$^{\star}$ & 0.17 (0.07)$^{\star}$ \\
    Machine Learning Literacy & 0.14 (0.14) & -0.62 (0.40) & 0.10 (0.11) \\
    Experience with XAI User Studies: Yes & -0.10 (0.21) & 1.35 (0.59)$^{\star}$ & -0.15 (0.16) \\ 
    Education + Age & \multicolumn{1}{c}{---} & -0.07 (0.05) & \multicolumn{1}{c}{---} \\
    Education + Male & \multicolumn{1}{c}{---}  & 0.08 (0.14) &  \multicolumn{1}{c}{---} \\
    Education + STEM & \multicolumn{1}{c}{---} & -0.56 (0.19)$^{\star\star}$ & \multicolumn{1}{c}{---} \\   
    Education + Machine Learning Literacy & \multicolumn{1}{c}{---} & 0.17 (0.10) & \multicolumn{1}{c}{---} \\ 
    Education + Experience with XAI User Studies & \multicolumn{1}{c}{---} & -0.35 (0.16)$^{\star}$ & \multicolumn{1}{c}{---} \\
    \midrule
    $R^2$ & \multicolumn{1}{c}{0.117} & \multicolumn{1}{c}{0.111} & \multicolumn{1}{c}{0.070} \\
    Adjusted $R^2$ & \multicolumn{1}{c}{0.089} & \multicolumn{1}{c}{0.058} & \multicolumn{1}{c}{0.040} \\  
    \bottomrule
    \end{tabular}
    \label{tab:individual_diff}
\end{table}

\begin{figure}
    \centering
    \begin{subfigure}[b]{0.42\linewidth}
    \centering
     \includegraphics[width=\linewidth]{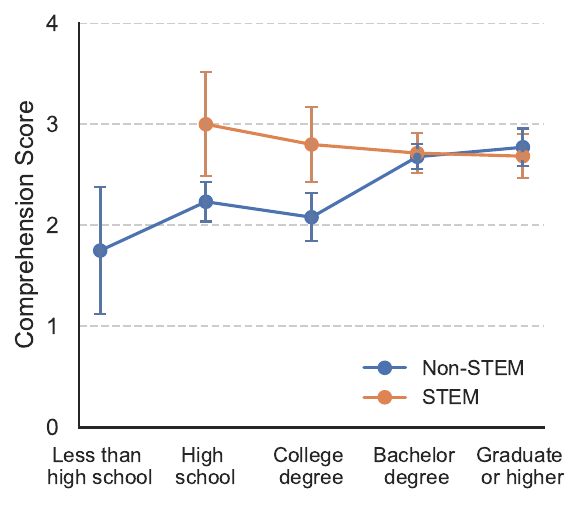}
    \caption{Education vs.\ STEM background.}\label{fig:edu_interaction_0}
    \end{subfigure}
    \hspace{2em}
    \begin{subfigure}[b]{.4\linewidth}
    \centering
     \includegraphics[width=\linewidth]{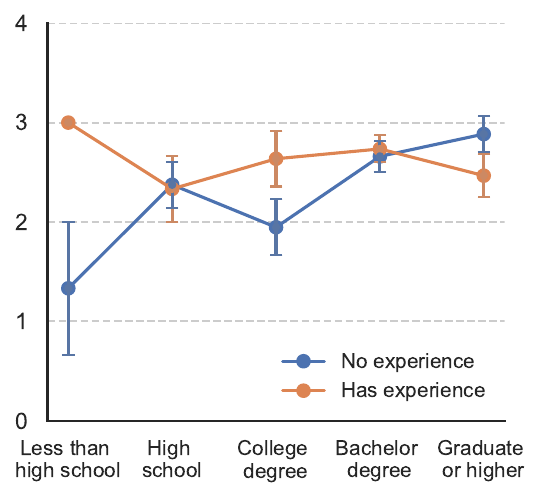}
    \caption{Education vs.\ user study experience.}\label{fig:edu_interaction_1}
    \end{subfigure}
    \caption{
    Average score for the four questions about \emph{explicated information} stratified by education level and (\subref{fig:edu_interaction_0}) STEM background or (\subref{fig:edu_interaction_1}) XAI user study experience. On average, participants who had STEM background or experience with XAI user studies achieved \emph{lower} comprehension scores as their education level increased. On the contrary, participants without STEM background or XAI user study experience attained \emph{higher} scores as their education level increased. Error bars indicate standard error.}
    \label{fig:edu_interaction}
\end{figure}

\subsection{Individual Difference}

We further explore the effect of individual characteristics on explanation comprehension. The major personal characteristics we consider in our user study include age, gender, STEM background, education level, machine learning literacy and experience with XAI user studies. We treat age, ML literacy and education as ordinal data; gender as categorical data; and the remaining variables as binary data. Our sample consists of five non-binary participants, who we exclude from our analysis in this section due to their limited representation in our sample. Table~\ref{tab:individual_diff} describes the generalised linear models that we use to analyse the effect of personal characteristics on the participants' scores for the comprehension questions about explicated and unspecified information as well as scores for all the questions. (Recall that each participant answered four questions about explicated information and four questions about unspecified information during the experiment, thus the score for each set of questions is in the 0--4 range and the total score is in the 0--8 range.)   
 
Models~5 and 7 -- detailed in Table~\ref{tab:individual_diff} -- describe main effects of individual characteristics on \emph{overall comprehension} and \emph{comprehension of unspecified information} respectively. Additional models were constructed to explore interaction effects in this setting, but since  
no significant interaction effects were observed we do not include these results. Model~6 explores main effects of individual characteristics on \emph{comprehension of explicated information}; it identified a significant main effect of education level on this type of comprehension. We further explore interaction effects between education level and the remaining five personal characteristics on comprehension of explicated information; the results are displayed in Table~\ref{tab:individual_diff} as Model~6.

Model~5 shows that the participants' education level was positively correlated with overall comprehension, with the average comprehension score increasing as the attained education level becomes higher ($\mathit{coef}$ $=$ 0.33, $p$ $<$ 0.01). Models~6 and 7 confirm the significantly positive correlation between education level and comprehension for explicated ($\mathit{coef}$ $=$ 0.50, $p$ $<$ 0.05) and unspecified ($\mathit{coef}$ $=$ 0.17, $p$ $<$ 0.05) information. 
Model~6 shows that technical background had a significantly positive effect on the comprehension of \emph{explicated information} ($\text{coef}$ $=$ 2.31, $p$ $<$ 0.01), with participants who had STEM background achieving higher score ($\mu$ $=$ 2.74, $\mathit{SD}$ $=$ 1.00) than those without it ($\mu$ $=$ 2.46, $\mathit{SD}$ $=$ 1.04). Experience with XAI user studies was also positively correlated with comprehension of \emph{explicated information} ($\mathit{coef}$ $=$ 1.35, $p$ $<$ 0.05), with veteran participants achieving higher scores ($\mu$ $=$ 2.50, $\mathit{SD}$ $=$ 0.88) than their inexperienced peers ($\mu$ $=$ 2.51, $\mathit{SD}$ $=$ 1.12). 

Model~6 also highlights two other significant interaction effects on user comprehension of \emph{explicated information}: between education and STEM background ($\mathit{coef}$ $=$ -0.56, $p$ $<$ 0.01), and between education and XAI user study experience ($\mathit{coef}$ $=$ -0.35, $p$ $<$ 0.05). We visualise these two interaction effects in Figure~\ref{fig:edu_interaction}. On average, participants with STEM background achieved lower comprehension scores as their education level increased. On the contrary, participants without STEM background attained higher scores as their education level increased. Similarly, participants who have never before participated in XAI user studies scored higher when their education level was higher; this trend is reversed -- worse performance was associated with higher education levels -- for participants who interacted with XAI user studies before. We discuss the implications of this finding in Section~\ref{sec:dis-ind-char}. 

Finally, we analyse the relationship between participants' graph literacy and their XAI comprehension. Pearson correlation tests show that the correlation between the graph literacy score and correct answers to the questions about \emph{explicated information} was significant, with the average comprehension score decreasing as the participants' graph literacy increased ($\mathit{coef}$ $=$ -0.20, $p$ $<$ 0.01). We do not observe any significant relationship between graph literacy and overall comprehension score or comprehension score for unspecified information alone. 

\begin{figure}[t]
    \centering
    \begin{subfigure}[b]{0.42\linewidth}
    \centering
     \includegraphics[width=\linewidth]{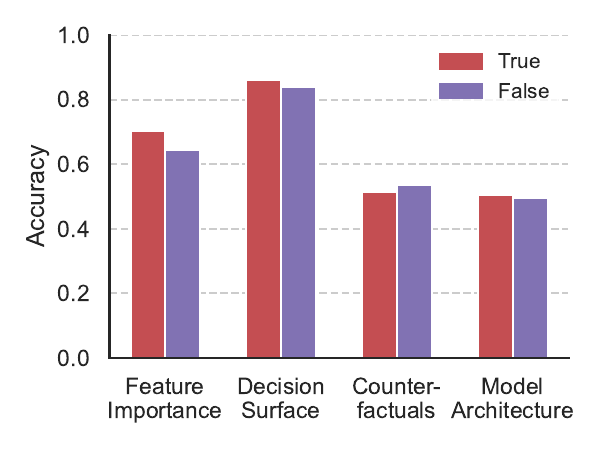}
    \caption{Accuracy of the ``true'' and ``false'' explicated-information questions.}\label{fig:tf_acc}
    \end{subfigure}
    \hspace{2em}
    \begin{subfigure}[b]{.42\linewidth}
    \centering
     \includegraphics[width=\linewidth]{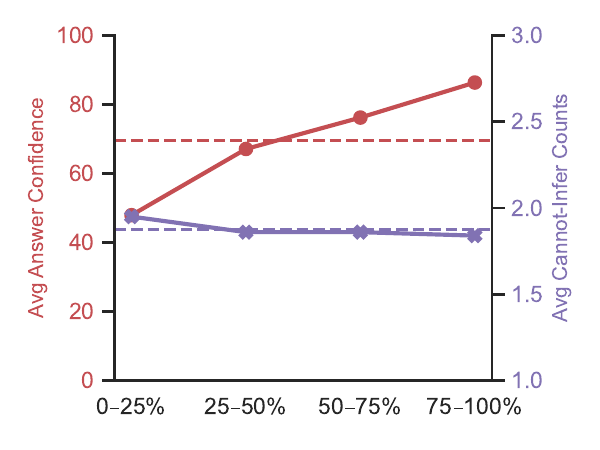}
    \caption{Frequency of selecting ``cannot infer'' as the answer among different participant groups.}\label{fig:ct_conf}
    \end{subfigure}
    \caption{
    Overview of our participants' engagement analysis. Panel~(\subref{fig:tf_acc}) shows the accuracy of the answers to the ``true'' or ``false'' assertions grouped by explanation type. Both the ``true'' and ``false'' assertions are gauging user comprehension of explicated information. For each explanation, the difference in answer accuracy to the ``true'' and ``false'' questions is insignificant. Panel~(\subref{fig:ct_conf}) shows the frequency of selecting the ``cannot infer'' option as the answer to the comprehension questions. We group participants into quartiles based on their average answer confidence, with each group having an equal number of participants. The first quartile (0--25\%) consists of participants with the lowest confidence and the fourth quartile (75--100\%) with the highest confidence. The red dotted line indicates the average answer confidence across all the participants; the purple dotted line indicates the average number of times a participant chooses ``cannot infer'' as their answer out of the eight comprehension questions. The figure shows that the participants whose confidence was low opted for the ``cannot infer'' option more frequently than average, however the difference is not significant.}
    \label{fig:response_analysis}
\end{figure}

\subsection{Participant Engagement Analysis}\label{sec:rigour}
To verify the validity and robustness of our results, we conduct additional data analysis to study the participants' engagement with the user study. Given that the ``true'' assertions regarding counterfactual explainability are negatively framed, which may lead to the concern that they confuse participants, it is important to ensure a comparable accuracy between the ``true'' and ``false'' assertions across the explanations. We confirm this in Figure~\ref{fig:tf_acc}, which shows that the participants who received the ``true'' assertion questions exhibited similar performance to those who received the ``false'' assertions. This suggests that the negatively framed assertions had a negligible impact on user comprehension.

We further explore if the participants actively engaged in more cognitively demanding tasks (e.g.,~detecting unspecified information) or whether instead they frequently opted for the ``cannot infer'' option as their default (passive) answer. Figure~\ref{fig:ct_conf} shows that our participants chose the ``cannot infer'' option 1.875 times on average out of the eight questions they received. For participants with low confidence in their answers, they tended to choose ``cannot infer'' more frequently than average (1.95 times). However, the difference in frequency is small among the participants with different confidence levels. This suggests that the participants engaged actively in detecting unspecified information instead of defaulting to the ``cannot infer'' option.

Lastly, we analyse the time that the participants spent on each explanation. We find that the participants spent more time attempting to understand global explanations, which communicate more information and involve more features: 3.16~mins on model architecture and 3.11~mins on feature importance. On the other hand, the participants spent less time on local explanations, which only present information about two features; specifically, 3.08~mins on decision surface visualisation and 2.94~mins on counterfactual explainability. This suggests that the participants invested more effort in understanding more complex explanations as expected.

\section{Discussion}\label{sec:discussion}

In this paper, we set out to understand whether users can correctly interpret
factual insights conveyed by machine learning explanations and, at the same
time, be aware of their inherent limitations. Current literature emphasises
that bespoke explanatory mechanisms are needed to facilitate understanding of
data-driven predictive systems as transparency by itself cannot guarantee
intelligibility~\citep{huysmans2011empirical,lage2019human,jesus2021can,sokol2023reasonable}.   
Our findings complement this perspective by demonstrating that the 
\textbf{comprehension of explanations is a double-edged sword as highly intelligible explanations are likely to be misinterpreted despite their inherent truthfulness}. 
In this section, we discuss our results with a particular focus on user comprehension and misinterpretation of explanations. Then, we suggest better explanation modelling strategies to reduce the prevalence of such undesired effects.

\subsection{Explanation Misinterpretation}%

Our results demonstrate that our \textbf{participants were good at correctly interpreting the information explicated by explanations, but struggled to identify the information that remained unspecified}.  
For example, feature importance informs explainees how much a model relies on each feature. Figure~\ref{fig:overall_compare_0} shows that participants achieved a high comprehension level for the information explicated by feature importance compared to other explanation types. Nonetheless, participants also (incorrectly) inferred local feature influence -- i.e.,~whether a feature contributes positively or negatively to a particular prediction -- from such explanations despite this information being unavailable.  
We also find that users tend to identify the features used in the explanations
based on decision surface visualisation and counterfactual explainability as
the most important, even though such explanations do not (explicitly) account
for this information. We hypothesise that participants may exhibit confirmation
bias -- a well-known cognitive error -- which in this case led the explainees
to assume that the feature affecting a particular prediction, as shown by the
explanation, must also be the most important
one~\citep{kruglanski1983bias,ghassemi2021false}. 
Our results additionally indicate that users tend to identify the feature used by the
root node of a decision tree as the most important one, which is not
necessarily correct given that this property is determined by an impurity
criterion rather than the tree structure; this finding aligns with the
phenomenon observed by~\citet{bell2022s}. 

In terms of specific explainability approaches, our results show that
participants were significantly more likely to understand explicated
information when it is communicated by feature importance and decision surface
visualisation rather than counterfactual explainability and model architecture.
A similar finding was reported by \citet{cheng2019explaining}, who showed that
white-box explanations (similar to our feature importance) increase explainees'
objective understanding.\footnote{Note that the definition of objective understanding
used by \citet{cheng2019explaining} corresponds to our notion of comprehension
of explicated information.} 
We additionally find that the former two explanation types were also more
likely to be misinterpreted, i.e.,~the participants were prone to deduce
unspecified information from them, evidenced by a low level of comprehension of
unspecified information shown in Figure~\ref{fig:overall_compare_0}. A similar
finding was observed by \citet{chromik2021think} who indicated that users
overestimate the understanding
they gain from local feature explanations because of the illusion of explanatory depth.
Counterfactual explainability and model architecture were less intelligible to participants in terms of explicated information, but the participants were also less likely to be misled by these explanations. To summarise, our results show that, among the four explanation types, \textbf{when the information explicated by an explanation was more comprehensible, unspecified information was also more likely to be misconstrued by the explainees}.

We also find that participants did not achieve a high level of comprehension of explicated information for model architecture when compared to the remaining three explanation types -- as shown in Figure~\ref{fig:overall_compare_0} -- even though this approach discloses the complete model structure, thus allows its in-vivo simulation.  
This aligns with the argument put forth by \citet{sokol2023reasonable} who
noted that transparency is a prerequisite for understanding, but this property
alone is insufficient for comprehensibility. Similarly, \citet{bell2022s}
argued that disclosing more information does not necessarily yield more
comprehension as the effectiveness of information processing is also crucial.
Furthermore, participants were less likely to correctly recognise the
unspecified information in explanations produced for decision trees and more
acute to it for logistic regression models. Given that the explicated
information of the explanations generated for both models was equally
comprehensible -- refer to Figures~\ref{fig:overall_compare_1} \& \ref{fig:overall_compare_2} -- this finding suggests that the explainees were
more likely to misinterpret explanations for predictors that were seemingly
easy to understand. This aligns with the argument described by \citet{bell2022s} that a
less complex predictor could be more misleading than a complicated one.
In summary, validating
user comprehension of explicated information when evaluating the intelligibility of user-centred explanations is \emph{insufficient}; it is equally critical to assess the degree of confusion and misinterpretation that XAI may lead to. 

\subsection{Over-generalisation and Over-confidence}

Through the objective assessment of comprehension, we find that participants did not know how much they knew; in other words, they could not correctly assess their state of knowledge. 
As shown in Figure~\ref{fig:tf-ct-comp}, having a high level of comprehension of explicated information did not stop users from over-generalising the explanations. We also incorporated subjective assessment results, and paired participants' subjective answer confidence with objective answer accuracy to evaluate whether the self-assessed confidence in their judgement exceeded the correctness of their comprehension. 
It is highly desirable for participants to correctly understand the factual insights conveyed by an explanation while also having high confidence in their comprehension. The opposite, however, should be avoided: participants should not overlook the limitations of an explanation and be confident in the misconstrued information. 

In this work, we find that \textbf{participants who interpreted explicated information correctly also possessed higher confidence in their correct interpretation than others who incorrectly interpreted explicated information} across all explanation types. This observation suggests that the explanations led to high confidence that is well calibrated for explicated information. In contrast, \textbf{participants became overconfident in their incorrect interpretation of the information unspecified by explanations}, especially so for feature importance and model architecture. 
This indicates that not only the limitations of these explanations were difficult to identify, but also that the information unspecified by these two explanation types appeared convincing enough for users to develop high confidence in their misinterpretation.  

Prior work on overconfidence noted that humans' confidence exhibits low levels
of calibration when individuals face difficult
tasks \citep{lichtenstein1977those,kruger1999unskilled}. In this work, we
observe a similar phenomenon where participants had a high degree of
overconfidence when working on a more difficult task, namely answering
questions about unspecified information, for which the correct response rate
was significantly lower as demonstrated by Figure~\ref{fig:overall_compare}. This
suggests that identifying the limitations of explanations is a more challenging
task than just attempting to understand explicated information, in which case
user confidence is more likely to be poorly calibrated.  
As \citet{darwin1888descent} noted decades ago, ``ignorance more frequently
begets confidence than does knowledge''. 
Our findings confirm that people who are ignorant of the limitations of an explanation are more overconfident than their peers who are more competent.

\subsection{User Perception}

In general, the \textbf{participants agreed that feature importance and
decision surface visualisation were easier to understand and sufficiently
detailed as compared to the other two explanation types}. The explainees
believed that counterfactual explainability was harder to understand and not
sufficiently detailed, despite this XAI approach being praised in the
literature for its appeal, simplicity and
brevity \citep{wachter2017counterfactual,karimi2022survey,sokol2023reasonable}.
This suggests that simple explanations do not necessarily provide the
information expected or needed by explainees, and the right balance between
information load and simplicity should be considered on a case-by-case basis.
Model architecture was perceived as hard to understand and insufficiently
detailed. This suggests that providing all information about an ML model does
not necessarily guarantee that users find an explanation sufficiently
detailed; one possible reason is the details being unintelligible due to the
explanation complexity, which can be seen in Figure~\ref{fig:overall_compare_0} as
low comprehension of explicated information. 

The participants' perception of whether an explanation is easy to understand corresponded to their comprehension level of explicated information. As shown in Figure~\ref{fig:sub_comp}, the participants' subjective agreement with easiness of understanding was high for decision surface visualisation but low for model architecture. By additionally considering Figure~\ref{fig:overall_compare_0}, we find that the explanations for which people achieved high comprehension of explicated information but low comprehension of unspecified information were found to be universally easier to understand. In summary, \textbf{whenever an explanation was believed to be ``easy to understand'', it was objectively more intelligible but at the same time also more misleading}.

\subsection{Individual Characteristics}\label{sec:dis-ind-char}

Correct and truthful understanding of an explanation relies on human
reasoning~\citep{sokol2021explainability,sokol2023reasonable}, but reasoning
capacity varies between individuals, which in turn affects the degree of
understanding. Human factors are therefore of paramount importance when
studying user comprehension. 
Our results indicate that the \textbf{participants with higher educational attainment developed a more accurate understanding of both the information that an explanation explicated and the information that remained unspecified}.  
Those with lower education levels were more prone to develop various misinterpretations.  
This finding agrees with the well-known Dunning--Kruger effect, which states
that those with limited knowledge in a domain not only reach incorrect
conclusions but their incompetence prevents them from realising their
mistakes~\citep{kruger1999unskilled}. Similarly, \citet{cheng2019explaining}
found that explainees' education level has a main effect on their understanding
of predictive algorithms. 

We also observe that \textbf{users with technical background developed a more
accurate interpretation of explicated information}, which confirms findings in
prior work~\citep{laato2022explain}. 
Furthermore, \textbf{participants who have taken part in XAI user studies
before performed better in grasping the information explicated by
explanations}. This suggests that ML explanations could be an effective
educational tool to increase user knowledge of AI-related
concepts~\citep{ng2021ai}. 
On the other hand, 
Human--Computer Interaction researchers studying explainability should be cautious of crowd workers with prior XAI user study experience, and ideally include their historical participation in such experiments as a confounding factor. Since our user study strictly followed the standard participant recruitment protocol popular in current XAI literature, our results offer a warning that this factor may influence the participants' performance. While the individuals involved in user studies might gain XAI knowledge from their exposure, this nuanced factor cannot be easily captured through demographic questionnaires employed in current study designs. Future work could also explore the nature of this type of experience.

\subsection{Explanation Complementarity}\label{sec:framework}%

Providing information that is incomplete, partial or oversimplified may
engender a false sense of knowledge and lead to
overconfidence~\citep{lackner2023intermediate}. 
Our work uncovered that explanations for which the users struggled to identify
their limitations and unspecified information were likely to be misinterpreted
(with poor confidence calibration). Such misleading explanations could also
possibly be misuse~\citep{parasuraman1997humans} and result in unwarranted
trust~\citep{jacovi2021formalizing}. 
Furthermore, according to \citet{leichtmann2023effects}, a high level of trust
in ML models should not be a goal in itself as excessive trust may lead to
judgement errors and other unintended consequences due to unjustified
overreliance. Therefore, when users' only interaction with predictive models is
through explanations, 
it is crucial to help them develop an appropriate, well-calibrated level of confidence and trust in their comprehension. 
To this end, we posit that practitioners should use explanations -- or combinations thereof that are complementary -- with well-understood limitations that are either known to the explainees, easy to identify or explicitly communicated. 

One solution, suggested by \citet{van2021effect}, is for researchers to
explicitly indicate the information that is unspecified by (explanatory)
artefacts shown to humans. However, we argue that simply disclosing the
limitations of an explanation may not suffice to engender understanding and
well-calibrated trust.
This is because numerous explanations are available, each with a different scope -- global, local or cohort -- and information content -- attribute importance, feature influence, counterfactual explainability, model architecture, and the like. 
It is thus impractical to identify and communicate all the pieces of information unspecified by a single explanation or their collection. 
Additionally, it remains to be seen whether explainees can understand and correctly apply such limitations and how to best communicate them. 

As one possible approach to address this challenge, we suggest to \textbf{consider the alignment of different explanations, thus identify the ones that provide complementary information, covering the shortcomings and limitations of each other}. Complementary explanations promise to offer non-overlapping information without overwhelming users. As an example, our participants were likely to infer feature importance from decision surface visualisation and counterfactual explainability; by accompanying such explanations with feature importance, we can prevent incorrect insights from developing. In this context, we can consider feature importance to be complementary to decision surface visualisation and counterfactual explainability since it minimises the chances of explainees inferring insights that remain unspecified. Future work could investigate if complementary explanations can mitigate this type of misinterpretations as compared to standalone explanations. 

\subsection{Limitations and Future Work}

We recognise several limitations of our work that should be considered when interpreting our results. (1)~We only investigated four common explainability approaches, looking at the degree to which participants can identify explanatory information and its limitations. Future work could evaluate other XAI techniques and identify explanation types that lead to a high level of comprehension for both explicated and unspecified information. (2)~As discussed in Section~\ref{sec:ml-model}, we only explored inherently interpretable and transparent models, setting aside complex AI models for which we could not ensure perfect explanation fidelity. Future work could extend our study to black-box models, accounting for the imperfect fidelity of their explanations.  
(3)~The scenario presented to our participants -- chronic disease diagnosis -- covered only the healthcare domain. 
Future work could explore other scenarios and application domains that span diverse levels of risk associated with data-driven decision-making. 

Additionally, (4)~we limited the design of our comprehension statements to quiz-style questions pertaining exclusively to information about data features. However, an explanation could provide multi-faceted information depending on its type and scope, hence users may deduce different kinds of unspecified information. Future work could thus expand the set of comprehension questions to cover more types of explicated and unspecified information. This approach would help to develop a comprehensive view of the information captured by each explanation and the limitations thereof. Such a framework could be used to better align explanations to provide complementary and reliable information. 
(5)~The assertions vary across explanations as they are designed to capture the specific information communicated by different explanations. Thus one may find that the structure and difficulty of the assertions differ. As we followed the framework described in Section~\ref{sec:framework} and analysed the impact of the assertion design in Section~\ref{sec:rigour}, we posit that the formulation of the assertions has a negligible impact on the study results. Future work could design a skeleton of assertion statements so that one can easily plug in different pieces of information to consistently formulate the assertion questions.
Finally, (6)~we note that our user study was limited to UK crowd workers and different demographics may produce different results.

\section{Conclusion\label{sec:conclusion}}

Providing explanations that are intelligible to the general public, such that the recipients do not misconstrue the information that they carry,
is of paramount importance. 
In this paper, we investigated whether common machine learning explanations provide intelligible insights and to what extent they are likely to be misinterpreted (when their limitations are overlooked). 
Specifically, we examined user comprehension of explicated and unspecified information for four representative explainability approaches applied to two popular types of predictive models: logistic regression and decision tree; this set-up allowed us to assess user perception of explanation informativeness and difficulty. 
We found that comprehension can be a double-edged sword as highly intelligible explanations are often misinterpreted, resulting in incorrect beliefs. 
In particular, we showed that feature importance and decision surface visualisation offer highly comprehensible information, but 
the explainees are ignorant of their limitations, thus misinterpret the insights that they provide. 
Further, our findings indicate that explanations perceived by the users as ``easy to understand'' are in fact objectively more comprehensible, but also more misleading. 
Accounting for the explainees' perception of the explicated and unspecified information can therefore
inspire better design of the explanation content and presentation format, such that it provides 
complementary insights that address each other's limitations or combat these shortcomings otherwise. 

\renewcommand{\acksname}{Acknowledgements}
\begin{acks}
This research was conducted by the ARC Centre of Excellence for Automated Decision-Making and Society (project number CE200100005), and funded by the Australian Government through the Australian Research Council.
Additional funding was provided by the Hasler Foundation (grant number 23082).
\end{acks}

\section*{Author Contribution}

\textbf{Yueqing Xuan}:
Writing -- original draft, 
Writing -- review \& editing, 
Methodology, 
Formal analysis, 
Conceptualization.
\textbf{Edward Small}: Methodology, Conceptualization.
\textbf{Kacper Sokol}: 
Writing -- original draft, 
Writing -- review \& editing, 
Validation, 
Methodology, 
Conceptualization, 
Supervision.
\textbf{Danula Hettiachchi}: Writing -- original draft, 
Validation, 
Methodology, 
Conceptualization.
\textbf{Mark Sanderson}: Validation, 
Methodology, 
Supervision, 
Funding acquisition.

\bibliographystyle{ACM-Reference-Format}
\bibliography{references}

\clearpage\newpage
\appendix

\section{Survey Flow}\label{app:flow}
\begin{figure}[h!]
    \centering
    \includegraphics[width=1.00\linewidth]{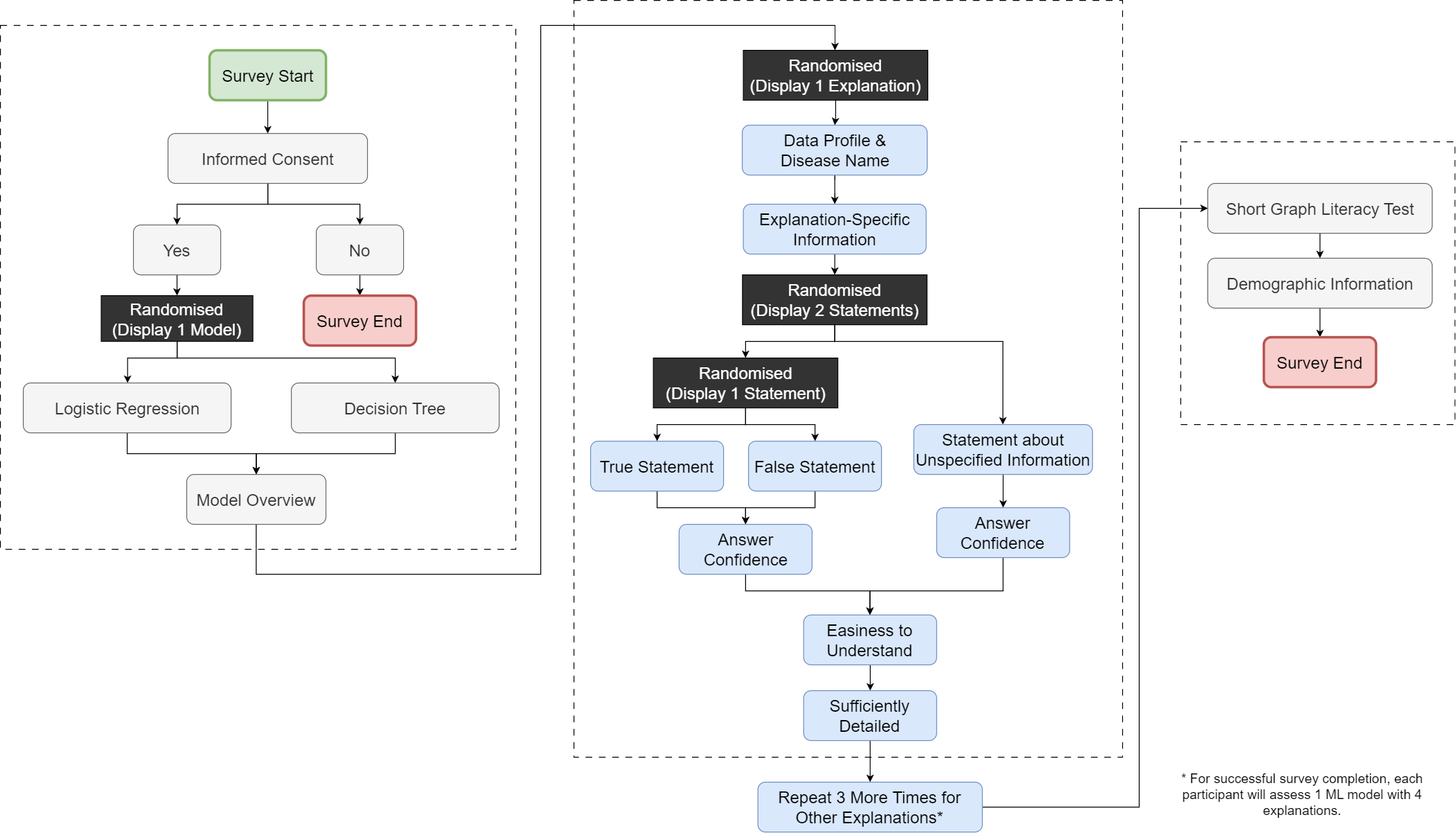}
    \caption{Diagram illustrating the flow of our survey.}%
    \label{fig:flow-chart-simple}
\end{figure}
\begin{figure}[t!]
    \centering
    \includegraphics[width=1.00\linewidth]{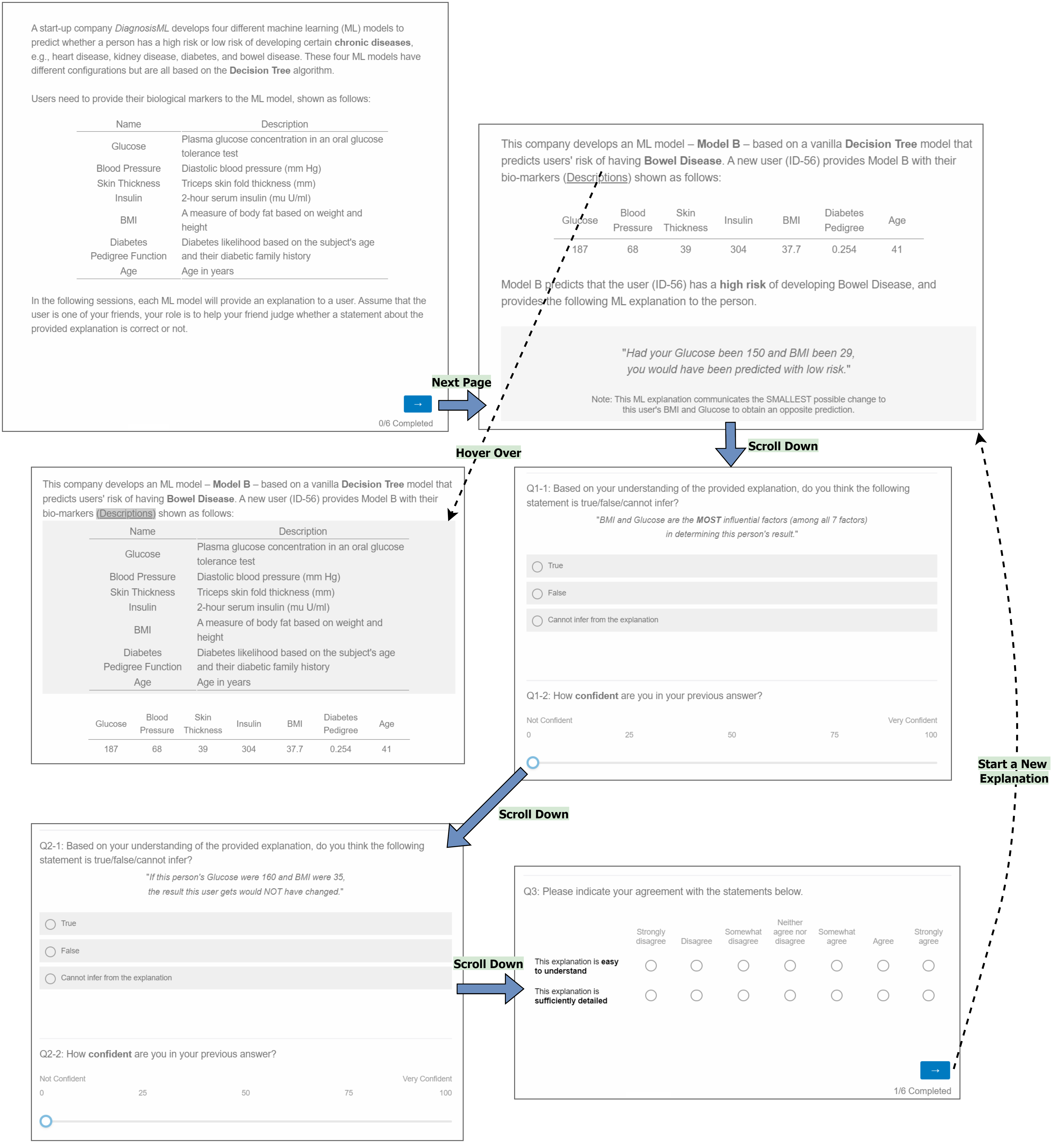}
    \caption{%
    Presentation of the task description and explanations in the survey. For brevity, we use only one explanation as an example.}%
    \label{fig:flow-chart-example}
\end{figure}

\section{Explanations and Comprehension Statements}
\label{app:exp}

\begin{table}[t!]
\small
\caption{Explanations and comprehension assertions for the \emph{logistic regression} model. The explanations are, from top to bottom, feature importance, decision surface visualisation, counterfactual explainability, and model architecture.}%
\begin{tabular}{@{}p{6cm}p{7cm}@{}}
\toprule
\multicolumn{1}{c}{Explanation} & \multicolumn{1}{c}{Comprehension Assertion} \\  \midrule
        \raisebox{-1.0\height}{
\includegraphics[width=5.925cm]{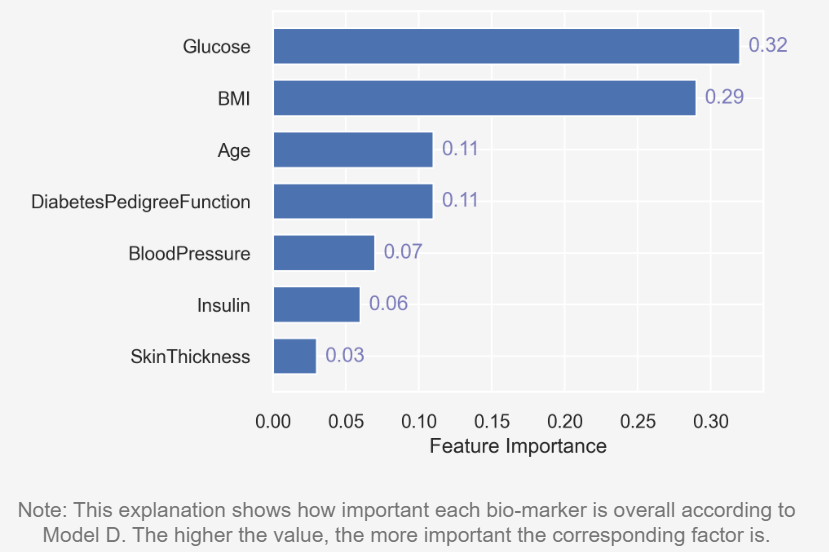}} &
\begin{description}[topsep=0pt]%
    \item [True] Model D uses all 7 biomarkers while making predictions, and Glucose is the most important factor for Model D.
    \item [False] For Model D, Age is more important than DiabetesPedegreeFunction
when predicting users' risk of diabetes.
    \item [Cannot Tell] According to Model D, increasing one's BMI and Insulin would increase the predicted risk of having Diabetes.
\end{description}\nointerlineskip\\
        \raisebox{-1.0\height}{
\includegraphics[width=5.925cm]{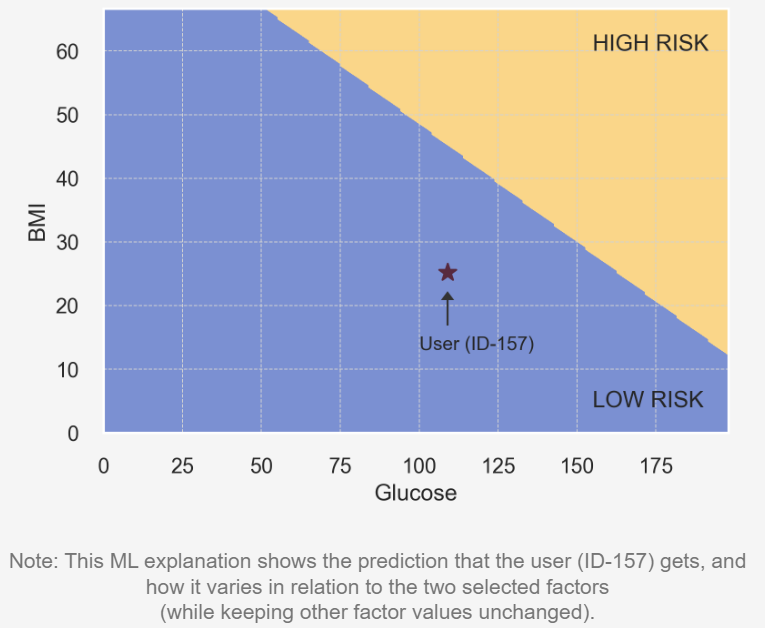}} &
\begin{description}[topsep=0pt]
\item [True] Assuming that all other biomarker values remain the same, increasing this person's Glucose to 150 and BMI to 40 will change their prediction from low risk to high risk.
 \item [False] Assuming that all other biomarker values remain the same (including BMI), increasing this person's Glucose to 135 will change their prediction from low risk to high risk.
 \item [Cannot Tell] BMI and Glucose are the MOST influential factors (among all 7 factors) in determining this person's result.
\end{description}\nointerlineskip\\
        \raisebox{-1.0\height}{
        \centering
\includegraphics[width=5.925cm]{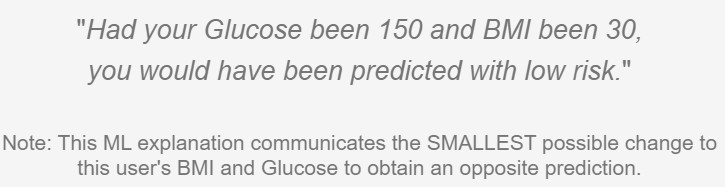}} &
\begin{description}[topsep=0pt]
    \item [True] If this person's Glucose were 160 and BMI were 35, the result this user gets would NOT have changed.
    \item [False] If this person's BMI were 30 while all the other factors remained unchanged (including Glucose), the result this user gets would have changed to low risk.
    \item [Cannot Tell] BMI and Glucose are the MOST influential factors (among all 7 factors) in determining this person's result.
\end{description}\nointerlineskip\\
        \raisebox{-1.0\height}{
\includegraphics[width=5.925cm]{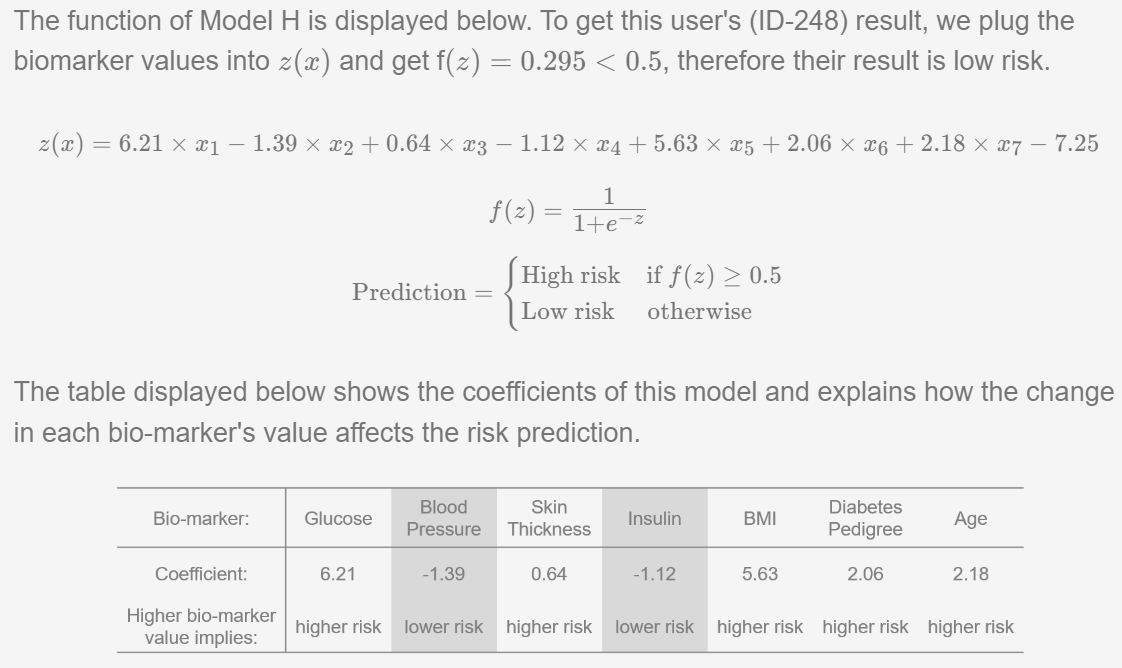}} 
 &
\begin{description}[topsep=0pt]
    \item [True] According to Model H, decreasing a person's Glucose and BMI will decrease their predicted risk.
    \item [False] According to Model H, decreasing a person's Glucose and BMI will increase their predicted risk.
    \item [Cannot Tell] According to Model H, increasing a person's Blood Pressure and BMI will decrease their predicted risk.
\end{description}\nointerlineskip\\
 \bottomrule
\end{tabular}
\label{tab:lr_questions}
\end{table}

\begin{table}[t]
\small
\caption{Explanations and comprehension assertions for the \emph{decision tree} model. The explanations are, from top to bottom, feature importance, decision surface visualisation, counterfactual explainability, and model architecture.}%
\begin{tabular}{@{}p{6cm}p{7cm}@{}}
\toprule
\multicolumn{1}{c}{Explanation} & \multicolumn{1}{c}{Comprehension Assertion} \\  \midrule
        \raisebox{-1.0\height}{
\includegraphics[width=5.925cm]{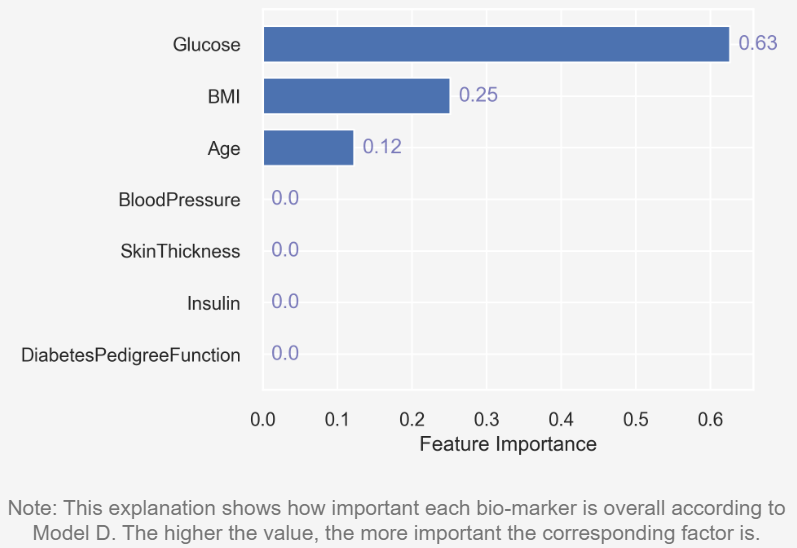}} &
\begin{description}[topsep=0pt]
    \item [True] Model D only uses 3 features to predict users' risks of diabetes,
and Insulin level does not affect the model's predictions.
\item [False] The level of Insulin influences Model D's prediction for this user (ID-24) and all other users.
\item [Cannot Tell] According to Model D, increasing one's BMI and Age will increase the predicted risk of having Diabetes.
\end{description}\nointerlineskip\\

        \raisebox{-1.0\height}{
\includegraphics[width=5.925cm]{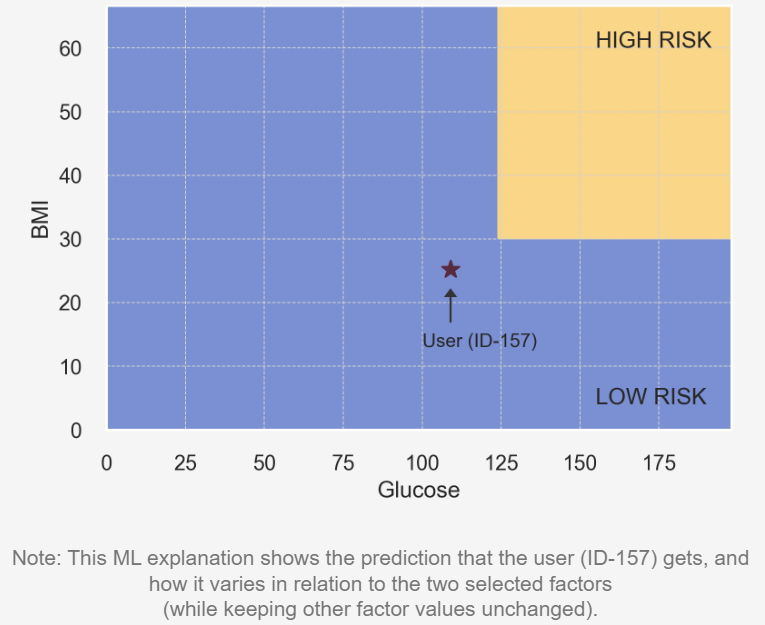}} &
\begin{description}[topsep=0pt]
\item [True] Assuming that all other biomarker values remain the same, increasing this person's Glucose to 135 and BMI to 36 will change their predicted result from low risk to high risk.
\item [False] Assuming that all other biomarker values remain the same (including BMI), increasing this person's Glucose to 135 will change their result from low risk to high risk.
\item [Cannot Tell] BMI and Glucose are the MOST influential factors (among all 7 factors) in determining this person's result.
\end{description}\nointerlineskip\\

        \raisebox{-1.0\height}{
\includegraphics[width=5.925cm]{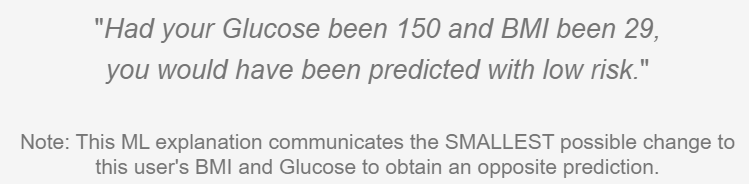}} &
\begin{description}[topsep=0pt]
\item [True] If this person's Glucose were 160 and BMI were 35, the result this user gets would NOT have changed.
\item [False] If this person's BMI were 29 while all the other factors remained unchanged (including Glucose), the result this user gets would have changed to low risk.
\item [Cannot Tell] BMI and Glucose are the MOST influential factors (among all 7 factors) in determining this person's result.
\end{description}\nointerlineskip\\

        \raisebox{-1.0\height}{
\includegraphics[width=5.925cm]{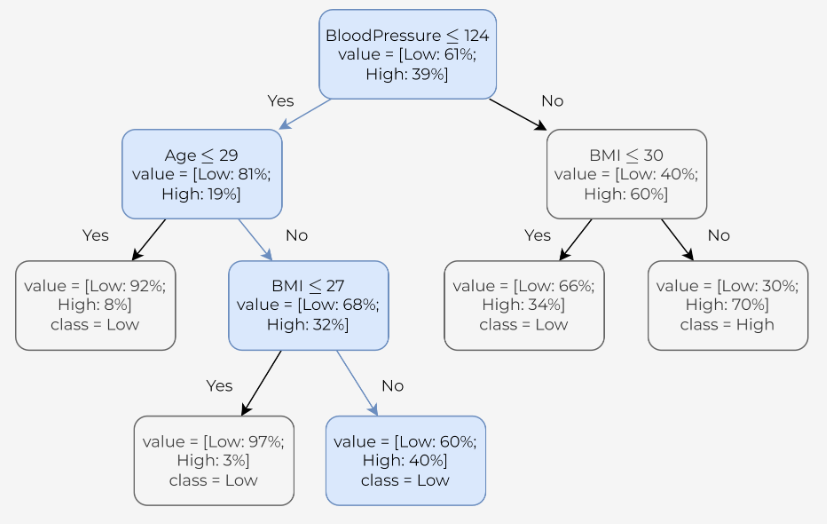}} 
 &
\begin{description}[topsep=0pt]
\item [True] Assuming that all other biomarker values remain the same, increasing this person's Blood Pressure to 130 will change their result from low risk to high risk.
\item [False] Assuming that all other biomarker values remain the same, increasing this person's BMI to 38 will change their result from low risk to high risk.
\item [Cannot Tell] Blood Pressure has the most impact (i.e., is the most important) on the prediction this user (ID-248) gets.
\end{description}\nointerlineskip\\
 \bottomrule
\end{tabular}
\label{tab:dt_questions}
\end{table}

\clearpage
\section{Outcomes of Additional Statistical Tests}\label{app:stats_results}

\begin{table}[h!]
    \centering
    \small
    \caption{$\mathcal{X}^2$ and significance indicators ($^{\star{}}$ for $p< 0.05$, $^{\star{}\star{}}$ for $p<0.01$ and $^{\star{}\star{}\star{}}$ for $p<0.001$) of McNemar's test.
    The results show that feature importance and decision surface visualisation are more likely to lead to accurate comprehension of \emph{explicated information} than counterfactual explainability and model architecture.}
    \begin{tabular}{rllll}
    \toprule
    & Feature importance & Decision surface & Counterfactual & Model architecture \\ \midrule
    Feature importance  & \multicolumn{1}{c}{---} & 18.28$^{\star\star\star}$ & 9.18$^{\star\star\star}$ & 14.41$^{\star\star\star}$ \\
    Decision surface   & \multicolumn{1}{c}{---} & \multicolumn{1}{c}{---} &44.47$^{\star\star\star}$ & 52.13$^{\star\star\star}$ \\
    \bottomrule
    \end{tabular}
    \label{tab:pairwise_comp}
\end{table}

\begin{figure}[h!]
    \centering
    \includegraphics[width=\linewidth]{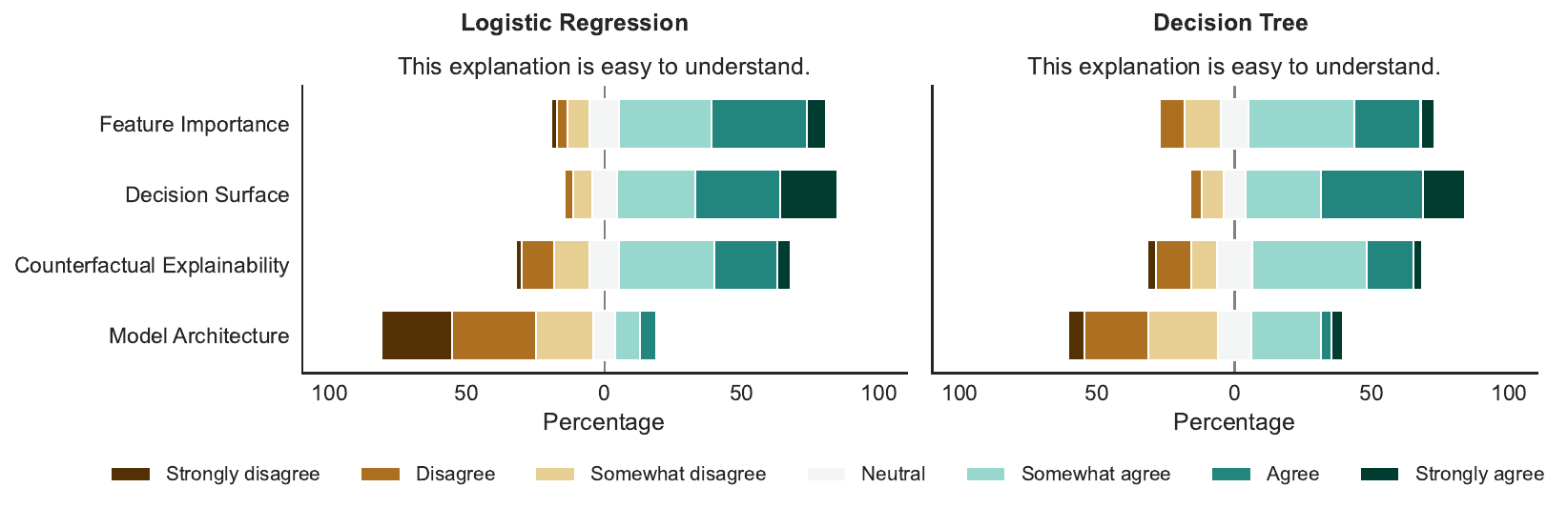}
    \caption{Overview of Likert scale responses to the ``this explanation is easy to understand'' question stratified by explanation type for logistic regression (left) and decision tree (right).}
    \label{fig:easy_per_model}
\end{figure}

\begin{figure}[h!]
    \centering
    \includegraphics[width=\linewidth]{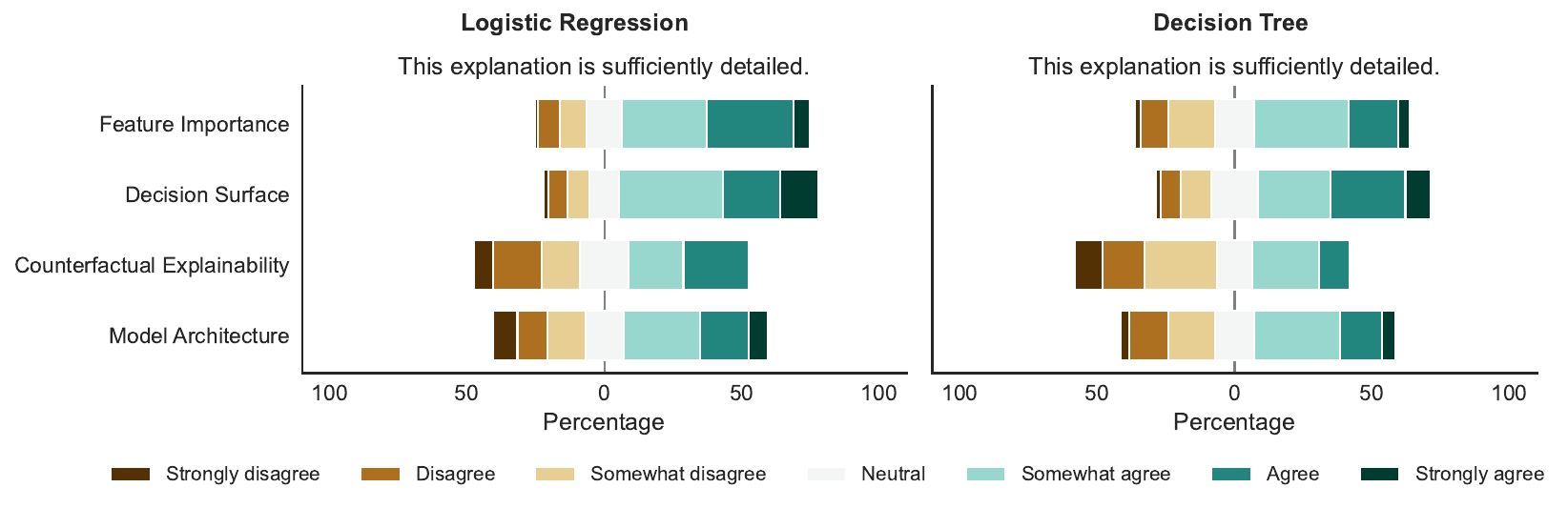}
    \caption{Overview of Likert scale responses to the ``this explanation is sufficiently detailed'' question stratified by explanation type for logistic regression (left) and decision tree (right).}
    \label{fig:detail_per_model}
\end{figure}

\begin{table}[h!]
    \centering
    \small
    \caption{Difference between the mean and significance indicator ($^{\star{}}$ for $p< 0.05$, $^{\star{}\star{}}$ for $p<0.01$ and $^{\star{}\star{}\star{}}$ for $p<0.001$) of
    Tukey's post-hoc paired test (the results are diagonally symmetrical). 
    The outcomes show comparative differences between participants' responses to the ``this explanation is easy to understand'' question for different explanation types.}
    \begin{tabular}{rllll}
    \toprule
    & \multicolumn{1}{c}{Feature importance} & \multicolumn{1}{c}{Decision surface} & \multicolumn{1}{c}{Counterfactual} & \multicolumn{1}{c}{Model architecture} \\ \midrule
    Feature importance & \multicolumn{1}{c}{---} & -0.475$^{\star\star}$& 0.400$^{\star}$ & 1.785$^{\star\star\star}$ \\
    Decision surface & \multicolumn{1}{c}{---} & \multicolumn{1}{c}{---} & 0.875$^{\star\star\star}$ & 2.260$^{\star\star\star}$ \\
    Counterfactual & \multicolumn{1}{c}{---} & \multicolumn{1}{c}{---} & \multicolumn{1}{c}{---} & 1.385$^{\star\star\star}$ \\
    \bottomrule
    \end{tabular}
    \label{tab:easy_ttest}
\end{table}

\begin{table}[h!]
    \centering
    \small
    \caption{Difference between the mean and significance indicator ($^{\star{}}$ for $p< 0.05$, $^{\star{}\star{}}$ for $p<0.01$ and $^{\star{}\star{}\star{}}$ for $p<0.001$) of
    Tukey's post-hoc paired test (the results are diagonally symmetrical). 
    The outcomes show comparative differences between participants' responses to the ``this explanation is sufficiently detailed'' question for different explanation types.}
    \begin{tabular}{@{}rllll}
    \toprule
    & \multicolumn{1}{c}{Feature importance} & \multicolumn{1}{c}{Decision surface} & \multicolumn{1}{c}{Counterfactual} & \multicolumn{1}{c}{Model architecture} \\ \midrule
    Feature importance & \multicolumn{1}{c}{---} & -0.235 & 0.830$^{\star\star\star}$ & 0.395$^{\star}$ \\
    Decision surface & \multicolumn{1}{c}{---} & \multicolumn{1}{c}{---} & 1.065$^{\star\star\star}$ & 0.630$^{\star\star\star}$ \\
    Counterfactual & \multicolumn{1}{c}{---} & \multicolumn{1}{c}{---} & \multicolumn{1}{c}{---} & -0.435$^{\star}$ \\  
    \bottomrule
    \end{tabular}
    \label{tab:detail_ttest}
\end{table}

\section{Additional Analysis of Priming Effect}\label{app:priming}

\begin{figure}[h!]
    \centering
    \begin{subfigure}[t]{0.41\linewidth}
    \centering
     \includegraphics[height=.72\linewidth,trim={0 -0.5cm 0 0},clip]
     {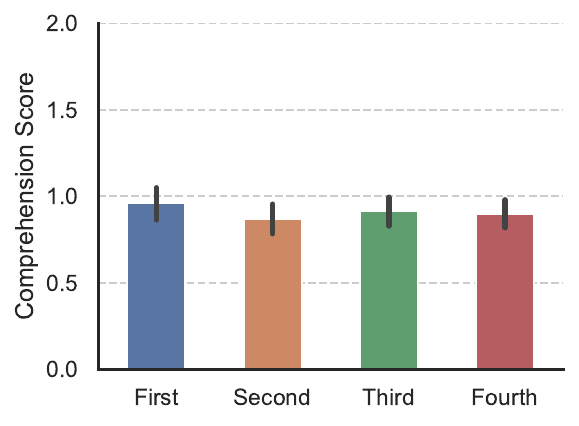}
    \caption{Random order.}\label{fig:priming-effect_0}
    \end{subfigure}
    \hspace{2em}
    \begin{subfigure}[t]{0.4\linewidth}
    \centering
     \includegraphics[height=.75\linewidth]{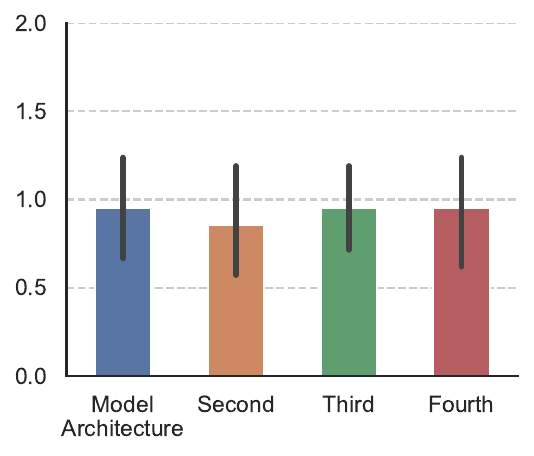}
    \caption{Model architecture shown first.}\label{fig:priming-effect_1}
    \end{subfigure}
    \caption{
    Average number of questions answered correctly per explanation. We fully randomise the order of explanations in each survey to remove the impact of the priming effect. Panel~(\subref{fig:priming-effect_0}) shows that the average performance for each explanation is comparable regardless of the order in which explanations were provided. Panel~(\subref{fig:priming-effect_1}) demonstrates that for those users who were shown \emph{model architecture} as their first explanation their performance on subsequent tasks did not improve consistently. Error bars indicate the 95\% confidence interval.}
    \label{fig:priming-effect}
\end{figure}

As discussed in Section~\ref{sec:procedure}, we designed our user study as to eliminate the priming effect. In this appendix, we provide further evidence that the ordering effect did not play a role in the participants' performance as the result of our design choice. The performance results shown in Figure~\ref{fig:priming-effect} suggest that participants did not gain an additional advantage when they were exposed to the \emph{model architecture} explanation at the beginning, which could have been the case given that it is a global explanation that contains complete information about the model. More precisely, their performance in subsequent tasks remained stable -- see Figure~\ref{fig:priming-effect_1} -- and followed a similar pattern to the one found when the order of explanations was fully randomised -- see Figure~\ref{fig:priming-effect_0}.

\end{document}